\newcommand{\beq}{\begin{eqnarray}}
\newcommand{\eeq}{\end{eqnarray}}
\newcommand{\ti}{\tilde{t}}
\newcommand{\ra}{\tilde{r}}
\newcommand{\vit}{\tilde{\theta}}
\newcommand{\vip}{\tilde{\phi}}
\newcommand{\curv}{1-\frac{R_S}{\tilde{r}}}
\newcommand{\curvconst}{1-\frac{R_S}{\tilde{r}_0}}
\newcommand{\crosst}{\frac{R_S}{\tilde{r}}\,c}
\newcommand{\radvelf}{R_S\left(\frac{1}{\tilde{r}}-\frac{1}{\tilde{r}_0}\right)}
\newcommand{\hcurv}{1-\left(\frac{\tilde{r}}{R_H}\right)^2}
\newcommand{\hcurvconst}{1-\left(\frac{\tilde{r}_0}{R_H}\right)^2}
\newcommand{\hcrosst}{\left(\frac{\tilde{r}}{R_H}\right)^2c}

\documentclass[prb]{revtex4}
\usepackage{amsmath}

\usepackage{graphicx}

\begin{document}

\title{A river model of space}

\author{Simen Braeck}
\email{Simen.Brack@hioa.no}
\author{\O yvind Gr\o n}
\email{Oyvind.Gron@hioa.no}
\affiliation{Oslo and Akershus University College of Applied Sciences, Faculty of Engineering, P.O. Box 4 St. Olavs Plass, N-0130 Oslo, Norway}

\begin{abstract}
Within the theory of general relativity gravitational phenomena are usually attributed to the curvature
of four-dimensional spacetime. In this context we are often confronted with the question of how the concept
of ordinary physical three-dimensional space fits into this picture. In this work we present a simple and
intuitive model of space for both the Schwarzschild spacetime and the de Sitter spacetime in which physical
space is defined as a specified set of freely moving reference particles. Using a combination of orthonormal
basis fields and the usual formalism in a coordinate basis we calculate the physical velocity field of these
reference particles. Thus we obtain a vivid description of space in which space behaves like a river flowing
radially toward the singularity in the Schwarzschild spacetime and radially toward infinity in the de Sitter
spacetime. We also consider the effect of the river of space upon light rays and material particles and show
that the river model of space provides an intuitive explanation for the behavior of light and particles at
and beyond the event horizons associated with these spacetimes.
\end{abstract}

\maketitle

\section{Introduction\label{sec:intro}}

In teaching the theory of relativity we often meet the question: What is space? It is here understood that the question
is concerned with ordinary three-space and not the four-dimensional spacetime. The first part of the answer is to make
clear that space is a theory dependent concept. The second is to try to explain what we mean by 'space' according to the
general theory of relativity.

One definition is to say that space is a set of simultaneous events. Even if this is an essential part of what we mean by
space, this definition is not sufficient to give us a picture of space which makes us understand for example why light
can not be emitted from the horizon of a black hole. Also, we must demand from the properties of space that they make us
understand that special relativity is valid locally even in curved spacetime. Furthermore, the definition should capture the
phenomenon of inertial dragging, which will be considered in a later work.

In four-dimensional spacetime the concept of ordinary three-space is not uniquely defined. Due to the relativity
of simultaneity the separation of spacetime into space and time depends upon the motion of the reference
particles that define the three-space. For example, the spherical symmetry of space outside the center of a static,
spherically symmetric distribution of mass implies that it is mathematically convenient to choose a family of
reference particles which reflect this spatial symmetry as well as the static property of this spacetime, which leads to
the famous Schwarzschild metric. We shall call the class of coordinates that are comoving with a rigid reference frame
defined by a family of observers who remain at rest outside the mass distribution for Schwarzschild coordinates. Each one
of these observers is equipped with a \emph{coordinate} clock synchronized with the coordinate clocks of all the other
observers. These coordinate clocks thus define the time coordinate $t$ of the Schwarzschild coordinates. Setting
$t=\mbox{const.}$ then defines the geometry of three-space corresponding to this particular frame of reference.

However, although it often may be mathematically convenient to choose coordinates in which the three-space is closely
adapted to the symmetries, for example of the static character, of the particular spacetime being studied, this procedure
does not necessarily lead to the most natural definition of three-space from a physical point of view. For instance,
the rigid frame of reference associated with the Schwarzschild coordinates ceases to exist at and inside the horizon
of a black hole even though the Schwarzschild spacetime is perfectly regular in this region, and physical particles can
fall through the horizon. Accordingly, it might be more natural in this case to associate `physical' space with a
reference frame defined by the set of reference particles which fall freely from infinity and right through the horizon.
In this context we also note that freely moving material particles single out a preferred set of curves in
spacetime~\cite[p. 8]{Wald}, namely those curves which all material particles \emph{independently of their nature} follow
in the absence of any nongravitational influences (i.e., forces). In this sense free particles may be
regarded as defining a set of `natural' motions in spacetime.

Thus, motivated by the discussion above, we shall here define the concept of `physical space' as a continuum
of freely moving reference particles with specified initial conditions: In a space with a localized mass
distribution, the reference particles are assumed to be released with zero velocity from a region far from the mass
distribution in which a particle remaining at rest with respect to the mass has vanishing four-acceleration. In a
homogeneous and isotropic universe, the only motion of the reference particles is that due to a universal change
of the distances between the particles. This may be described by a single scale factor and defines the Hubble
flow which obeys Hubble's law. One may show~\cite{Gron} that these reference particles move freely.

\begin{figure}
\begin{center}
\includegraphics[width=10.0cm]{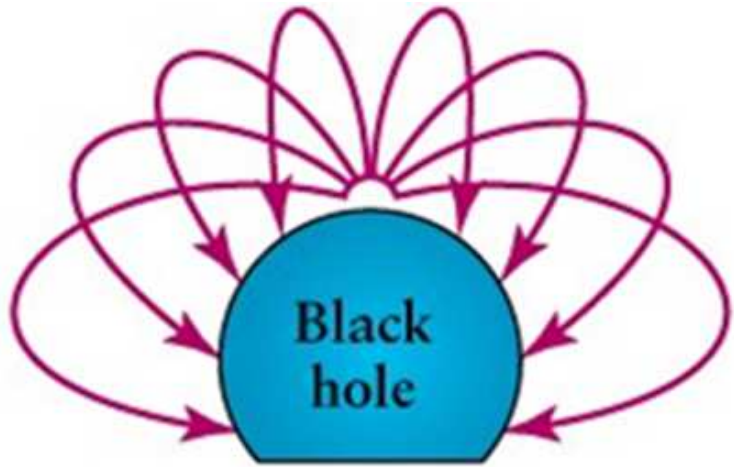}
\caption{Illustration of collapse to a black hole as presented in Ref.~\cite{Kaufmann}. There the author writes that
``as the curvature of spacetime around the star increases, light rays become increasingly deflected from their
usual straight paths. This deflection eventually becomes so severe that all light rays are bent back down
to the collapsing star's surface". The river model of space in the Schwarzschild spacetime presented in this work
demonstrates, however, that this `fountain' picture is misleading.}
\label{fig:fountain}
\end{center}
\end{figure}

In the present article we shall try to give a vivid impression of three-space as it appears in this `river model of space'.
In Sec.~\ref{sec:Schwarz} we establish the concept of the river model of space by considering the three-space of the
Schwarzschild spacetime. We shall see that this model makes it clear why the `fountain picture' in
Fig.~\ref{fig:fountain}, which has been presented in order to illustrate that light cannot escape from a black
hole~\cite{Kaufmann}, is misleading.

Our formulation of a river model of space in Schwarzschild spacetime is inspired by a recent version of the same
idea due to Hamilton and Lisle~\cite{Hamilton} who developed a river model of non-rotating black holes by expressing
the Schwarzschild metric in the so-called Gullstrand-Painlev\'{e} coordinates. In that model space itself is beautifully
pictured as a river flowing through a \emph{flat} background while objects moving \emph{in} the river move according
to the rules of special relativity. However, some of the basic features of this model contain elements of which the
description depends on the authors' particular choice of coordinates. For example, the velocity of the
inflowing river of space corresponds to the radial component of the coordinate velocity $dr/dt_{ff}$, where $r$ is
the radial coordinate and $t_{ff}$ is both the coordinate time and the proper time for a particle falling freely
in the radial direction. (This coordinate velocity does have an \emph{indirect} physical interpretation, however,
because the radial component of the four-velocity of a particle in radial free fall equals the Newtonian escape
velocity) Also, as the authors themselves recognize, the flat background ``has no physically observable meaning",
but is merely ``a fictitious construct that emerges from the mathematics".

In our version of the river model of space, the `background' relative to which the river of space
flows inward is defined by the physical reference frame associated with the stationary observers outside
the localized mass. In this way the velocity of the river of space in our model represents an invariant
physical quantity outside the Schwarzschild horizon which is independent of the choice of spacetime coordinates
and which, in principle, can be objectively measured. The advantage of this formulation of the model is
its immediate and intuitive connection to physical observations. Moreover, our concept of the river of space can easily
be generalized to many other spacetimes. This simple and intuitive formulation has a price, however, in that
it requires a somewhat unconventional interpretation of the `background' at and inside the
black hole horizon. Nevertheless, it will be shown that, in the case of the space outside a non-rotating black hole,
our formulation of the river model yields essentially the same picture as in `the river model of black holes'.

In Sec.~\ref{deSitter} we extend the river model of space by considering the three-space of the de Sitter spacetime.
The de Sitter spacetime represents a bridge between the static spherically symmetric spacetimes and the expanding
universe models. While the river of space flows towards the central mass in the Schwarzschild spacetime, it flows
outwards in the de Sitter spacetime. The `inertial river' is described in the same way as in the Friedmann universe
models by transforming from the original static coordinate system to a frame where the reference particles are the
local inertial frames.

With the river model of space we get a picture of a dynamical space in the spirit of the general theory of relativity.

\section{The river of space in Schwarzschild spacetime\label{sec:Schwarz}}

The Schwarzschild spacetime describes the geometry of the simplest example of a black hole as well as the empty
spacetime outside non-rotating stars. Our aim in this section is to establish the picture of a river model of space
in this spacetime in which `space' is defined as a set of material reference particles having vanishing
velocity at infinity and moving freely in the radial direction. The reference particles may be thought of as
so-called test particles~\cite{Hartle} having masses so small that they themselves do not influence the curvature
of spacetime. `Freely moving' means that the reference particles are not influenced by any (non-gravitational) forces.

In order to obtain the physical velocity field of the river of space, we begin by examining the radial trajectories of
freely falling particles in the Schwarzschild spacetime. Then we extend the analysis of the river of space to the
region inside the Schwarzschild horizon. In this connection we introduce a set of imaginary stationary observers
at and inside the horizon which are not physically real, but which nevertheless turn out to be useful in an
intuitive comparison of Schwarzschild black holes to the analogous Newtonian case. As a result, the river of space
flows faster than the speed of light relative to the stationary observers inside the horizon and thereby drags
all physical particles and light rays toward the singularity in this region.

\subsection{Coordinate velocity of a particle in radial free motion\label{subsec:freefallcoordvel}}

As already mentioned in the introduction, gravitational phenomena can be ascribed to the effects of spacetime
curvature upon the motion of free particles. Accordingly, one would expect that the three-space of the Schwarzschild
geometry may be naturally defined in terms of a reference frame constituting of a set of freely falling particles.
Indeed, the Schwarzschild spacetime can be described in terms of so-called Lema\^{i}tre coordinates to which one
associates such a reference frame. However, this description of the Schwarzschild spacetime does not correspond to
the gravitational effects induced by a central mass as they will appear to an observer at rest on the Earth's surface.
Hence, we shall relegate this description of the Schwarzschild spacetime to the Appendix.

Then, in Schwarzschild coordinates $(t,r,\theta,\phi)$, the Schwarzschild geometry is described by
the line element
\beq
ds^2 = -\left(1-\frac{R_S}{r}\right)c^2dt^2 + \left(1-\frac{R_S}{r}\right)^{-1}dr^2
 + r^2d\Omega^2\,,
\label{eq:Schwmetric}
\eeq
where $R_S=2GM/c^2$ denotes the Schwarzschild radius and $d\Omega^2=d\theta^2+\sin^2\theta\,d\phi^2$ is the solid
angle element. It is well known, however, that the metric given in~(\ref{eq:Schwmetric})
exhibits a coordinate singularity at $r=R_S$. The Schwarzschild coordinates thus fail to properly cover the region $r\leq R_S$
of the Schwarzschild spacetime. Since, in establishing the river model of space, we are interested also in future--directed
particle paths in the region $r\leq R_S$, we must choose a system of coordinates that appropriately describe the Schwarzschild
spacetime in the entire region $r>0$. The simplest coordinates that fulfill this requirement are the so-called
Eddington-Finkelstein coordinates $(\ti,\ra,\vit,\vip)$. They are related to the Schwarzschild coordinates by the coordinate
transformations
\beq
t = \ti - \frac{R_S}{c}\ln\left(\frac{r}{R_S}-1\right),\, r = \ra,\, \theta = \vit,\, \phi = \vip\,.
\label{eq:transf}
\eeq
This is an internal coordinate transformation~\cite[p. 79]{Gron} between different coordinates co-moving with the same
reference frame. By an internal transformation we mean a transformation for which the spatial coordinates of the new
coordinate system depends only on the spatial coordinates of the old system of coordinates.
Using these transformations in the line element in Eq.~(\ref{eq:Schwmetric}) the form of the line element in Eddington-Finkelstein
coordinates is calculated to be
\beq
ds^2 = -\left(\curv\right)c^2d\ti^2 + 2\crosst\, d\ti d\ra + \left(1+\frac{R_S}{\ra}\right)d\ra^2
+ \ra^2d\tilde{\Omega}^2\,.
\label{eq:edfmetric}
\eeq
This line element exhibits no coordinate singularity at $\ra=R_S$ and hence is suitable for describing particle paths in the
entire region $\ra>0$.

Note that the three-space of a static reference frame in the Schwarzschild spacetime is not obtained by putting
$\ti=\mbox{constant}$. In general the line element of a three-space orthogonal to the world lines of the reference
particles in a frame is given by~\cite{Gron}
\beq
dl^2 = \gamma_{ij}dx^{i}dx^{j}\,,
\label{eq.orthspatlinel}
\eeq
where
\beq
\gamma_{ij} = g_{ij} - \frac{g_{i0}g_{j0}}{g_{00}}
\label{eq:orthspatmetric}
\eeq
are the components of the spatial metrical tensor. The quantities $\gamma_{ij}$ transform like tensor components, and
the spatial line element is invariant under internal coordinate transformations.

With the line element in~(\ref{eq:edfmetric}),
\beq
g_{\ti\ti} = -\left(\curv\right)c^2,\, g_{\ra\ti} = \frac{R_S}{\ra}\,c,\, g_{\ra\ra} = 1+\frac{R_S}{\ra}
\label{eq:spatedfink}
\eeq
giving
\beq
\gamma_{\ra\ra} = \frac{1}{\curv}
\label{eq:spatmetradcomp}
\eeq
and
\beq
dl^2 = \frac{d\ra^2}{\curv} + \ra^2d\tilde{\Omega}^2\,,
\label{eq:spatlineledfink}
\eeq
in accordance with the spatial line element obtained by putting $dt=0$ in the line element~(\ref{eq:Schwmetric}).

Consider now a reference particle falling freely along the radial direction for which $d\vit=d\vip=0$, and assume that the particle
is released with zero velocity at the radial coordinate position $\ra=\ra_0$ at the coordinate time $\ti=0$.
The Lagrangian of the freely falling particle may then be written
\beq
L = \frac{1}{2}g_{\mu\nu}\dot{x}^{\mu}\dot{x}^{\nu}=-\frac{1}{2}\left(\curv\right)c^2\dot{\ti}^2
+ \crosst \dot{\ti}\dot{\ra} + \frac{1}{2}\left(1+\frac{R_S}{\ra}\right)\dot{\ra}^2\,,
\label{eq:SchwLag}
\eeq
where the dots denote the derivatives with respect to the proper time $\tau_{ff}$ of the freely falling particle. Since the Lagrangian
is independent of $\ti$, the momentum conjugate to the time--coordinate $p_{\ti}$ is a constant. Hence, $\dot{\ti}$ and $\dot{\ra}$
must satisfy the equation
\beq
p_{\ti} = \frac{\partial L}{\partial\dot{\ti}}=-\left(\curv\right)c^2\dot{\ti}+\crosst\dot{\ra} \,.
\label{eq:edfconstmotion}
\eeq
From the four-velocity identity $g_{\mu\nu}\dot{x}^{\mu}\dot{x}^{\nu}=-c^2$ we obtain the second equation that
$\dot{\ti}$ and $\dot{\ra}$ must satisfy as
\beq
p_{\ti}\dot{\ti} + \crosst\dot{\ti}\dot{\ra} + \left(1+\frac{R_S}{\ra}\right)\dot{\ra}^2
=-c^2\,.
\label{eq:fourvelid}
\eeq
Inserting the initial condition $\dot{\ra}\left|_{\ra=\ra_0}\right.=0$ in Eqs.~(\ref{eq:edfconstmotion}) and (\ref{eq:fourvelid}),
the constant $p_t$ is calculated to be
\beq
p_t = -\sqrt{\curvconst}\,c^2\,.
\label{eq:constofmotion}
\eeq
Obtaining $\dot{\ti}$ from Eq.~(\ref{eq:edfconstmotion}) and substituting the result in Eq.~(\ref{eq:fourvelid}) yields the
equation
\beq
-\left(p_{\ti} + \crosst\dot{\ra}\right)\frac{p_{\ti}-\crosst\dot{\ra}}{\left(\curv\right)c^2}
+ \left(1+\frac{R_S}{\ra}\right)\dot{\ra}^2 = -c^2\,,
\label{eq:fvelidconst}
\eeq
or
\beq
\dot{\ra} = -\sqrt{\radvelf}\,c\,,
\label{eq:fourradvel}
\eeq
where the negative sign has been chosen because the particle is falling inwards toward smaller $\ra$. The rate of the
coordinate time with respect to the proper time of the particle is
\beq
\dot{\ti} = \frac{\sqrt{\curvconst}-\frac{R_S}{\ra}\sqrt{\radvelf}}{\curv}\,.
\label{eq:fourtimvel}
\eeq
Finally, combining Eqs.~(\ref{eq:fourradvel}) and (\ref{eq:fourtimvel}), we obtain the radial coordinate velocity of a freely falling
particle as
\beq
\frac{d\ra}{d\ti} = - \frac{\left(\curv\right)\sqrt{\radvelf}}{\sqrt{\curvconst}-\frac{R_S}{\ra}\sqrt{\radvelf}}\,c\,.
\label{eq:radcoordvel}
\eeq
The river of space is defined by those freely falling particles that are initially at rest at infinity. Hence, the coordinate
velocity of the river of space is
\beq
\left(\frac{d\ra}{d\ti}\right)_{\mbox{space}} = \lim_{\ra_0\rightarrow\infty} \frac{d\ra}{d\ti}
= -\frac{\left(\ra-R_S\right)\sqrt{R_S}}{\ra^{3/2}-R_S^{3/2}}\,c\,.
\label{eq:spacecoordvel}
\eeq
In particular, using L'Hopital's rule, we find that the coordinate velocity of the river of space at the horizon $\ra=R_S$ is
$\left(d\ra/d\ti\right)_{\mbox{space}}=-(2/3)\,c$.

In our discussion of the Schwarzschild geometry, we will also be interested in the motion of light rays. Light rays move along null
geodesics for which $ds^2=0$. The coordinate velocity of light $\left(d\ra/d\ti\right)_l$ is therefore given by
\beq
\left(1+\frac{R_S}{\ra}\right)\left(\frac{d\ra}{d\ti}\right)_l^2 + 2\crosst \left(\frac{d\ra}{d\ti}\right)_l
- \left(1-\frac{R_S}{\ra}\right)c^2 = 0\,,
\label{eq:nullgeodesic}
\eeq
with solutions
\beq
\left(\frac{d\ra}{d\ti}\right)_{l+} = \frac{\ra-R_S}{\ra+R_S}\,c\ ,\ \left(\frac{d\ra}{d\ti}\right)_{l-} = -c\,
\label{eq:lightcoordvel}
\eeq
for outwards and inwards moving light rays, respectively. Accordingly, at the horizon of a black hole, the coordinate
velocities of light are $\left(d\ra/d\ti\right)_{l+}\left|_{\ra=R_S}\right.=0$ and $\left(d\ra/d\ti\right)_{\ra=R_S}=-c$.

\subsection{The physical velocity field of the river of space in the region $\ra>R_S$\label{subsec:riverexterior}}

The radial component of the coordinate velocity~(\ref{eq:radcoordvel}) does not, however, define a physical quantity that an observer
would measure in his local laboratory. Indeed, $d\ra/d\ti$ is the radial component of the velocity given with respect to the
coordinate basis vectors $\mathbf{e}_{\tilde{\mu}}$ which are generally neither unit vectors nor orthogonal. On the other hand, observers with
their associated local laboratories measure components of vector quantities with respect to their comoving orthonormal basis vectors.
For this reason we shall next introduce a field of orthonormal basis vectors corresponding to the local laboratories of stationary
physical observers in the Schwarzschild spacetime, and then calculate the radial velocity field of a freely falling particle
as measured by these stationary observers.

To construct a set of orthonormal basis vectors corresponding to a stationary observer, we first note that the timelike unit basis
vector $\mathbf{e}^s_{\hat{\tau}}$ associated with the stationary observer is identical to the observer's four-velocity
$\mathbf{u}_s$. This identification is directly carried over from special relativity to general relativity due to the equivalence
principle. Using that $d\ra=d\vit=d\vip=0$ for stationary observers, we thus obtain
\beq
\mathbf{e}^s_{\hat{\tau}} = \mathbf{u}_s = \frac{dx^{\tilde{\mu}}}{d\tau_s}\,\mathbf{e}_{\tilde{\mu}}
=\frac{1}{\sqrt{\curv}}\,\mathbf{e}_{\tilde{t}}\,,
\label{eq:statimeorto}
\eeq
where $\tau_s$ denotes the proper time of the stationary observer and the last equality follows directly from the line
element~(\ref{eq:edfmetric}) because $ds^2=-c^2d\tau_s^2$ for the time like path of the observer. Next, we let two of the three
spacelike unit basis vectors be aligned with the $\vit$-- and $\vip$--directions. Accordingly,
\beq
\mathbf{e}^s_{\hat{\theta}} =  \frac{1}{\tilde{r}}\,\mathbf{e}_{\vit}\,,\,
\mathbf{e}^s_{\hat{\phi}} = \frac{1}{\tilde{r}\sin\vit}\,\mathbf{e}_{\vip}\,.
\label{eq:stavinkorto}
\eeq
It is worth noting here that the three coordinate basis vectors $\mathbf{e}_{\tilde{t}}$, $\mathbf{e}_{\vit}$ and
$\mathbf{e}_{\vip}$ associated with Eddington-Finkelstein coordinates are equal to the three coordinate basis vectors
$\mathbf{e}_{t}$, $\mathbf{e}_{\theta}$ and $\mathbf{e}_{\phi}$ associated with Schwarzschild coordinates, respectively.
This may readily be confirmed by using the standard transformation rule
$\mathbf{e}_\mu=\left(\partial\tilde{x}^{\nu}/\partial x^{\mu}\right)\mathbf{e}_{\tilde{\nu}}$, where the transformation
matrix $\partial\tilde{x}^{\nu}/\partial x^{\mu}$ is given by Eq.~(\ref{eq:transf}). The spacelike unit basis vector
$\mathbf{e}^s_{\hat{r}}$, pointing in the increasing $r$--direction of the \emph{Schwarzschild} coordinates, may now be
found by noting that $\mathbf{e}^s_{\hat{r}}=\sqrt{1-R_S/r}\,\mathbf{e}_{r}$ and using the transformation rule
$\mathbf{e}_{r}=\left(\partial\tilde{x}^{\nu}/\partial r\right)\mathbf{e}_{\tilde{\nu}}$, giving
\beq
\mathbf{e}^s_{\hat{r}} = \frac{R_S}{\tilde{r}\sqrt{\curv}\,c}\mathbf{e}_{\tilde{t}} + \sqrt{\curv}\,\mathbf{e}_{\tilde{r}}\,.
\label{eq:staradorto}
\eeq
This result could also have been found without invoking the Schwarzschild coordinates by writing
$\mathbf{e}^s_{\hat{r}}=a_s^{\ti}\mathbf{e}_{\ti}+a_s^{\ra}\mathbf{e}_{\tilde{r}}$ and solving the two equations
$\mathbf{e}^s_{\hat{r}}\cdot\mathbf{e}^s_{\hat{\tau}}=0$ and $\mathbf{e}^s_{\hat{r}}\cdot\mathbf{e}^s_{\hat{r}}=1$ for
$a_s^{\ti}$ and $a_s^{\ra}$.
The set of mutually orthonormal basis vectors
$\{\mathbf{e}^s_{\hat{\tau}},\mathbf{e}^s_{\hat{r}},\mathbf{e}^s_{\hat{\theta}},\mathbf{e}^s_{\hat{\phi}}\}$
satisfy the requirements for an orthonormal basis, i.e.,
\beq
\mathbf{e}^s_{\hat{\mu}}\cdot\mathbf{e}^s_{\hat{\nu}} = \eta_{\hat{\mu}\hat{\nu}}\,,
\label{eq:reqorthobasis}
\eeq
where $\eta_{\hat{\mu}\hat{\nu}}$ denotes the Minkowski metric.

Using the same technique as M\"{u}ller~\cite{Muller}, we may now calculate the radial velocity field of a freely falling particle
by examining it's 4-velocity $\mathbf{u}_{ff}$. In the Eddington-Finkelstein coordinates we obtain
\beq
\mathbf{u}_{ff} = \frac{dx^{\tilde{\mu}}}{d\tau_{ff}}\,\mathbf{e}_{\tilde{\mu}}
= \dot{\ti}\,\mathbf{e}_{\ti}+\dot{\ra}\,\mathbf{e}_{\ra}
= \left(\mathbf{e}_{\ti}+\frac{d\ra}{d\ti}\,\mathbf{e}_{\ra}\right)\dot{\ti}\,,
\label{eq:fourveledfink}
\eeq
where $\dot{\tilde{t}}$ and $d\tilde{r}/d\tilde{t}$ are given by Eqs.~(\ref{eq:fourtimvel}) and (\ref{eq:radcoordvel}), respectively.
The stationary observers, however, measure physical quantities in their proper reference frames~\cite{MTW}.
Accordingly, let $(\tau_s,\hat{x}_s^r,\hat{x}_s^\theta,\hat{x}_s^\phi)$ denote the coordinates in the proper reference frame such that
the coordinate axes of the rectangular grid $(\hat{x}_s^r,\hat{x}_s^\theta,\hat{x}_s^\phi$) are aligned with the directions
of the orthonormal basis vectors $\{\mathbf{e}^s_{\hat{r}},\mathbf{e}^s_{\hat{\theta}},\mathbf{e}^s_{\hat{\phi}}\}$. Then,
in analogy with calculations of velocities within the special theory of relativity, the particle's coordinate velocity
$d\hat{x}_s^r/d\tau_{s}$ in an orthonormal basis is equivalent to the particle's \emph{physical} velocity as measured by the
stationary observers. In these coordinates the 4-velocity of the freely falling particle is given by
\beq
\mathbf{u}_{ff} &=& \frac{dx_s^{\hat{\mu}}}{d\tau_{ff}}\,\mathbf{e}^s_{\hat{\mu}}
= \left(\mathbf{e}^s_{\hat{\tau}}+\frac{d\hat{x}_s^r}{d\tau_s}\,\mathbf{e}^s_{\hat{r}}\right)\frac{d\tau_s}{d\tau_{ff}}\, \nonumber \\
&=&\left[\frac{1}{\sqrt{\curv}}\,\mathbf{e}_{\tilde{t}}
+ \frac{d\hat{x}_s^r}{d\tau_s}\left(\frac{R_S}{\tilde{r}\sqrt{\curv}\,c}\,\mathbf{e}_{\tilde{t}}
+\sqrt{\curv}\,\mathbf{e}_{\tilde{r}}\right)\right]\frac{d\tau_s}{d\tau_{ff}}\,.
\label{eq:fourvelortho}
\eeq
Since the expressions in Eqs.~(\ref{eq:fourveledfink}) and (\ref{eq:fourvelortho})
must be equal, this gives the two equations
\beq
\dot{\tilde{t}} = \frac{1}{\sqrt{\curv}}\left(1+\frac{R_S}{rc}\frac{d\hat{x}_s^r}{d\tau_{s}}\right)\frac{d\tau_s}{d\tau_{ff}},\;\ \;
\frac{d\tilde{r}}{d\tilde{t}}\,\dot{\tilde{t}} = \sqrt{\curv}\frac{d\hat{x}_s^r}{d\tau_{s}}\frac{d\tau_s}{d\tau_{ff}}\,.
\label{eq:fourvelequal}
\eeq
Eliminating $d\tau_s/d\tau_{ff}$ from these equations we find the general relation between the radial coordinate velocity
$d\ra/d\ti$ in the Eddington-Finkelstein system and the radial physical velocity $d\hat{x}_s^r/d\tau_{s}$ as measured by an
observer at rest in the reference frame,
\beq
\frac{d\ra}{d\ti} = \frac{\left(\curv\right)\frac{d\hat{x}_s^r}{d\tau_s}}{1+\frac{R_S}{\ra c}\frac{d\hat{x}_s^r}{d\tau_s}}\,,
\label{eq:gencoordortho}
\eeq
or
\beq
\frac{d\hat{x}_s^r}{d\tau_{s}} = \frac{\frac{d\ra}{d\ti}}{1-\frac{R_S}{\ra}\left(1+\frac{1}{c}\frac{d\ra}{d\ti}\right)}\,.
\label{eq:genorthocoord}
\eeq
Equation~(\ref{eq:gencoordortho}) reveals some interesting properties of the initial velocity of particles emitted by a stationary
observer at $\ra$. Firstly, if the observer emits light rays, $d\hat{x}_s^r/d\tau_s=\pm c$, one recovers Eq.~(\ref{eq:lightcoordvel}),
showing that although the physical velocity of light is isotropic in the local laboratory of the stationary observer, the coordinate
velocity of light in the Eddington-Finkelstein system is anisotropic. We also note that the coordinate velocity of light rays, which
are outgoing in the stationary observer's proper reference frame, vanish as $\ra\rightarrow R_S$. This analytical treatment thus shows
that the fountain picture of Fig.~\ref{fig:fountain} is misleading. As will be explicitly demonstrated in Sec.~\ref{subsec:rivermodel},
the anisotropic property of the coordinate velocity of light and its behavior at the Schwarzschild radius has a very intuitive
interpretation within the river model of space. Thus, we shall see that the river model of space has the advantage of providing a
simple pictorial framework that can be invoked in order to explain why the fountain picture must be rejected.

Secondly, the behavior of particles emitted just outside the horizon of a black hole can be examined by considering particles
emitted with the same physical velocity $d\hat{x}_s^r/d\tau_s=\pm ac$ ($a\leq 1$) in the positive and negative $\ra$--direction. The
coordinate velocity of the particle moving outwards is
\beq
\left(\frac{d\ra}{d\ti}\right)_+ = \frac{\left(\curv\right)a}{1+\frac{R_S}{\ra}\,a}\,c\,,
\label{eq:statcoordvelout}
\eeq
and for the particle emitted inwards,
\beq
\left(\frac{d\ra}{d\ti}\right)_- = -\frac{\left(\curv\right)a}{1-\frac{R_S}{\ra}\,a}\,c\,.
\label{eq:statcoordvelinw}
\eeq
These formulae show that the Eddington-Finkelstein coordinate velocities of particles emitted outwards approach zero as $\ra\rightarrow R_S$
for all fixed values of $a$, whereas the coordinate velocities of particles emitted inwards approach zero in the limit $\ra\rightarrow R_S$
for a fixed $a<1$. In contrast, particles emitted inwards at the speed of light, i.e., $a=1$, have coordinate velocities
$\left(d\ra/d\ti\right)_-=-c$ for all values of $\ra>R_S$. These results hold when we consider fixed values of the parameter $a$
while varying $\ra$. In general, however, $a$ will be a function of the coordinate time $\ti$, as in the case of a freely
falling particle.

Then, to obtain the physical velocity of a freely falling particle as measured in the stationary observers' reference frame,
we substitute the result of Eq.~(\ref{eq:radcoordvel}) in Eq.~(\ref{eq:genorthocoord}), yielding
\beq
\frac{d\hat{x}_s^r}{d\tau_{s}} = -\sqrt{\frac{R_S\left(\ra_0-\ra\right)}{\ra\left(\ra_0-R_S\right)}}\,c\,.
\label{eq:orthonormvelocity}
\eeq
Note that $\ra$ is both the instantaneous radial position of the freely falling particle as well as the position
of the stationary observer measuring its velocity.

\begin{figure}[tbp]
\begin{center}
\includegraphics[width=10.0cm]{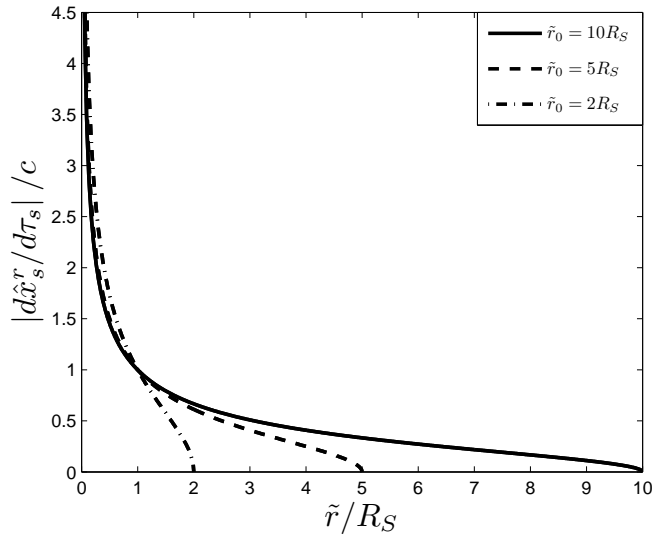}
\caption{Plots of the magnitude of the radial infall velocity $d\hat{x}_s^r/d\tau_{s}$ of a freely falling particle in
the Schwarzschild spacetime as measured by stationary observers in their proper reference frames. The different curves correspond
to different choices of the initial coordinate position $\ra_0$ of the freely falling particle. Note that, according to
Eq.~(\ref{eq:orthonormvelocity}), the freely falling particle appears to reach the speed of light at the Schwarzschild radius
$R_S$ (independently of its initial position $\ra_0$) and continues to increase its speed toward infinity as it approaches
the singularity at $\ra=0$.}
\label{fig:orthovel}
\end{center}
\end{figure}

Let us now analyze the expression in~(\ref{eq:orthonormvelocity}) for the region $\ra>R_S$ for which the concept of
stationary \emph{physical} observers is well-defined. The concept of stationary observers in the region $\ra\leq R_S$ will
be addressed in Sec.~\ref{subsec:riverinterior}. Two limiting cases are of particular interest. Firstly, as the freely falling
particle approaches the Schwarzschild radius, i.e., when $\ra\rightarrow R_S$, $d\hat{x}_s^r/d\tau_{s}\rightarrow -c$ for any
value of $\ra_0$. Thus, we may conclude that the speed of freely falling test particles in the Schwarzschild spacetime approach
the speed of light $c$ as they approach $R_S$ when measured with respect to a local reference frame at rest with the central
mass. These results are in concordance with the conclusions reached by several
authors~\cite{Cavalleri73,Cavalleri77,Cavalleri78,Janis73,Janis77,Crawford} who discussed the local velocity of a particle
falling freely from infinity. However, it should be emphasized that, although the freely falling particle may be said to
\emph{approach} the speed of light $c$ near the Schwarzschild radius $R_S$, it never actually reaches the speed of light
\emph{at} $R_S$ with respect to any \emph{physical} observer. This has been very clearly demonstrated by Crawford and
Tereno~\cite{Crawford}. The reason for this is that a physical observer, who must follow a timelike world line in spacetime,
can not stay at rest at $\ra=R_S$. At $\ra=R_S$ only photons, following a light like world line, can stay at rest
and all massive particles thus reach the speed of light only with respect to these stationary photons~\cite{Crawford}.
Figure~\ref{fig:orthovel} illustrates a few examples of radial infall velocities [Eq.~(\ref{eq:orthonormvelocity})] for
some particular choices of the initial position $\ra_0$.
It is worth noting here that, for reasons that will become apparent in the next section, the plots in Fig.~\ref{fig:orthovel}
of the infall velocities given by Eq.~(\ref{eq:orthonormvelocity}) have in fact been extended to the region $\ra\leq R_S$
even though physical observers cannot be stationary in this region.

Secondly, in the limit where the freely falling particle starts infinitely far away from the central mass, we find
\beq
\left(\frac{d\hat{x}_s^r}{d\tau_{s}}\right)_{\mbox{space}}=
\lim_{\ra_0\rightarrow\infty} \frac{d\hat{x}_s^r}{d\tau_{s}} = -\sqrt{\frac{R_S}{\ra}}\,c=-\sqrt{\frac{2GM}{\ra}}\,.
\label{eq:Newtonvel}
\eeq
Hence, the magnitude of $d\hat{x}_s^r/d\tau_{s}$ equals the Newtonian escape velocity in this limit. The velocity given
by Eq.~(\ref{eq:Newtonvel}) defines the velocity of the river of space.

\subsection{The river of space and imaginary stationary observers in the region $\ra\leq R_S$\label{subsec:riverinterior}}

The analysis presented in Sec.~\ref{subsec:riverexterior} showed that the local velocity of a freely falling particle with respect
to stationary observers becomes identical (in magnitude) to the Newtonian escape velocity [Eq.~(\ref{eq:Newtonvel})] in the
limit $\ra_0\rightarrow\infty$. Within the Newtonian theory of gravity, where gravity is conceived as a force, the Newtonian escape
velocity  corresponds to the velocity of a freely falling particle (from infinity) that would be observed by local stationary
observers who are at rest with respect to the central mass all along the freely falling particle's world line. In particular,
these stationary observers are supposed to correspond to real physical observers even in the region $\ra\leq R_S$. Then, according
to Eq.~(\ref{eq:Newtonvel}), freely falling observers will exceed the speed of light with respect to the stationary observers
in the region $\ra<R_S$. Within the Newtonian theory this intuitive picture poses no problem whatsoever because, in that theory,
there exists no upper limit for velocities. But how should we interpret Eq.~(\ref{eq:Newtonvel}) within a relativistic theory of
gravity, where such an upper barrier is represented by the speed of light $c$? After all, Eq.~(\ref{eq:Newtonvel}) was derived
entirely within the context of the general theory of relativity. In order to answer this question and to compare the Newtonian
picture of a freely falling particle outside a central mass to the analogous situation in the Schwarzschild spacetime, we shall
in the following discuss the concept of stationary observers inside the black hole horizon.

Then, inside the black hole horizon $\ra<R_S$, Eq.~(\ref{eq:orthonormvelocity}) predicts that the local speed of a freely falling
particle relative to the stationary observers exceeds the speed of light and approaches infinity as $\ra\rightarrow 0$.
Note that Eq.~(\ref{eq:reqorthobasis}) still holds in the region $\ra<R_S$, and the set of vectors given by
Eqs.~(\ref{eq:statimeorto})--(\ref{eq:staradorto}) therefore provide a well-defined orthonormal basis also inside the horizon.
We may now interpret this peculiar result by observing that, inside the horizon, stationary observers move along spacelike world lines,
i.e., world lines for which $ds^2>0$. This can be seen by putting $d\ra=d\vip=d\vit=0$ in the line element~(\ref{eq:edfmetric}),
which yields $ds^2=-\left(1-R_S/\ra\right)c^2d\ti^2$, and hence $ds^2>0$ for a stationary observer in the region $\ra<R_S$. In other
words, \emph{inside the black hole horizon stationary observers have properties similar to hypothetical faster-than-light particles
within the context of the special theory of relativity}~\cite{Sudarshan,Feinberg,Treumann}. Such hypothetical particles are
usually termed \emph{tachyons}\cite{Feinberg}, and we will follow that convention henceforth.

The analogy between the properties of stationary observers inside the black hole horizon and tachyons in special relativity
can be established as follows. In the coordinate basis the line element of a stationary observer is given by
\beq
ds^2 = -\left(1-\frac{R_S}{\ra}\right)c^2dt^2\,.
\label{eq:linelcoordb}
\eeq
The general line element in the orthonormal basis, valid on the accelerated observer's world line~\cite{MTW},
is given by
\beq
ds^2=-c^2d\tau_s^2+{d\hat{x}_s^r}^2+{d\hat{x}_s^\theta}^2+{d\hat{x}_s^\phi}^2\,,
\label{eq:linelorthob}
\eeq
and so for a stationary observer $ds^2=-c^2{d\tau_s}^2$. Equating the two expressions for the line element, we find that
\beq
d\tau_s = \sqrt{1-\frac{R_S}{\ra}}\,d\ti.
\label{eq:proptimestat}
\eeq
Thus, the proper time of a stationary observer becomes an imaginary quantity inside the black hole horizon.
Consider next a radial spatial displacement in the orthonormal basis for which
$d\tau_s=d\hat{x}_s^\theta=d\hat{x}_s^\phi=0$. Equation~(\ref{eq:linelorthob}) then yields $ds^2={d\hat{x}_s^r}^2$.
The same spatial displacement is given in the Schwarzschild coordinates by the conditions $d\theta=d\phi=0$ and
$dt=d\ti-R_S/(\ra(1-R_S/\ra))\,d\ra=0$, so that $d\ti=R_S/(\ra(1-R_S/\ra))\,d\ra$. Substituting this relation for $d\ti$
in the line element~(\ref{eq:edfmetric}) and using that $d\vit=d\vip=0$, we calculate the line element for a purely radial
displacement in Eddington-Finkelstein coordinates, yielding $ds^2=1/(1-R_S/\ra)\,d\ra^2$. As a result we obtain the proper
radial length measured by a stationary observer as
\beq
d\hat{x}^r=\frac{1}{\sqrt{1-\frac{R_S}{\ra}}}\,d\ra.
\label{eq:proplengthstat}
\eeq
Hence, inside the horizon $\ra<R_S$ proper lengths \emph{measured by stationary observers} are imaginary. Within the present
context it is also of interest to calculate the energy $E=-m_0\mathbf{u}_s\cdot\mathbf{u}_{ff}$ of a stationary observer with
respect to a freely falling observer. Utilizing Eqs.~(\ref{eq:statimeorto}) and (\ref{eq:fourveledfink}), and that
$\mathbf{e}_{\tilde{\mu}}\cdot\mathbf{e}_{\tilde{\nu}}=g_{\tilde{\mu}\tilde{\nu}}$, the energy is easily calculated to be
\beq
E = m_oc^2\sqrt{\frac{\curvconst}{\curv}}\,.
\label{eq:energy}
\eeq
Here $m_0$ denotes the rest mass of the stationary observer. For the energy to be a real number inside the horizon $\ra<R_S$
(assuming $\ra_0>R_S)$, the observer's rest mass $m_0$ must be imaginary. Hence, stationary observers at rest in the
region $\ra<R_S$ have been shown to exhibit several of the properties, as summarized by
Eqs.~(\ref{eq:proptimestat}), (\ref{eq:proplengthstat}) and (\ref{eq:energy}), that are usually associated with tachyons within
the context of special relativity~\cite{Sudarshan,Feinberg}.

An important remark concerning the concept of stationary observers inside the black hole horizon is appropriate here. Even though
the stationary observers inside the Schwarzschild horizon share many of the kinematical properties with tachyons, we do not
consider such observers to be physically real, but rather as a mathematical construct. For instance, the squared magnitude of
the four-acceleration $\mathbf{a}$ of such observers may be shown~\cite{Hartle} to be
\beq
\mathbf{a}\cdot\mathbf{a} = \frac{c^4R_S^2}{4\ra^4\left(\curv\right)}\,,
\label{eq:fouracc}
\eeq
and therefore $\mathbf{a}$ diverges at the horizon and becomes a timelike vector in the region $\ra<R_S$.
This does not mean, however, that this concept is not useful. On the contrary, as we will show below, this concept provides a
means to give an intuitive and rather direct comparison of black holes with the familiar Newtonian picture of free
particles outside a central mass. For these reasons, we shall henceforth refer to stationary observers inside the horizon as
"imaginary stationary observers".

The analysis presented above enables us to establish a picture of space in the Schwarzschild geometry as a river of
test particles freely falling relative to stationary observers who are at rest with respect to the spatial
Schwarzschild coordinates. The stationary observers have a timelike, lightlike and spacelike character outside, at
and inside the black hole horizon, respectively. The freely falling particles in turn represent inertial frames that define
space locally. Thus, the river of space, composed of the global collection of local inertial frames and which represents the
global space of the Schwarzschild spacetime, flows into the black hole according to Eq.~(\ref{eq:orthonormvelocity}). As
Fig.~\ref{fig:orthovel} shows, the river of space flows radially inward at speeds smaller than the speed of light outside
the horizon. Reaching the speed of light at the horizon, the river of space continues to fall at speeds exceeding the
speed of light with respect to the imaginary stationary observers inside the horizon and approaches infinite speed near
the singularity at $\ra=0$.

With this picture in mind we are now in a position to compare the Newtonian case of freely falling particles outside a
central point mass with the analogous situation in Schwarzschild spacetime. In the Newtonian case, particles in free fall
will simply continue to accelerate toward the central mass with respect to the stationary observers all along its world line,
approaching infinite speed relative to the stationary observers near the central mass. In the Newtonian theory both
the freely falling particles and the stationary observers are assumed to be real physical entities. As already mentioned,
this poses no problem in the Newtonian case since there exists no upper limit for increasing the relative velocity between
physical observers in this theory. In the Schwarzschild spacetime, however, the picture becomes radically different due
to the upper limit for relative velocities (i.e., the speed of light $c$) between physical observers. Because freely
falling observers in the Schwarzschild spacetime reach the speed of light relative to the imaginary stationary
observers at $\ra=R_S$, and further exceed the speed of light relative to the imaginary stationary observers for $\ra<R_S$,
\emph{all} real physical observers \emph{must} fall inward toward the central mass in the region $\ra\leq R_S$ in order not
to violate the upper limit for relative velocities. In other words, a physical observer with nonzero rest mass cannot stay
at rest at and inside the black hole horizon, but must continue his path in the direction of decreasing $\ra$ and eventually
hit the singularity at $\ra=0$.

\subsection{Coordinate velocity of a particle emitted by a freely falling observer\label{subsec:rivermodel}}

In Sec.~\ref{subsec:riverexterior} we utilized the stationary observers' local reference frames to calculate
the physical velocity field of the river of space. But, for the purpose of examining the effect of the river
of space upon the motion of light and particles, these local coordinate systems are not appropriate
because they do not properly capture the global properties of the Schwarzschild spacetime such as the
event horizon. Therefore, to further substantiate our interpretation of the effect of the river of space
upon particles given at the end of Sec.~\ref{subsec:riverinterior}, we now calculate the coordinate velocity
of a particle emitted from a freely falling observer in Eddington-Finkelstein coordinates.

To connect the three-velocity $v_e$ of the emitted particle as measured in the freely falling observer's laboratory
to the Eddington-Finkelstein coordinate velocity, we must first construct a set of orthonormal basis vectors
associated with an observer freely falling along the radial direction. The time like unit basis vector
$\mathbf{e}_{\hat{\tau}}^{ff}$ is simply given by the freely falling observers' four-velocity $\mathbf{u}_{ff}$.
Hence,
\beq
\mathbf{e}_{\hat{\tau}}^{ff} = \mathbf{u}_{ff}\,,
\label{eq:freetimeorto}
\eeq
where $\mathbf{u}_{ff}$ is given by Eq.~(\ref{eq:fourveledfink}). Once again we may choose two of the three spacelike
unit basis vectors to be aligned with the $\vit$-- and $\vip$--directions at the position $\ra=\ra_0$, where the
freely falling observer initially is at rest. Because of the spherical symmetry in this case, these unit basis
vectors will continue to be aligned with the $\vit$-- and $\vip$--directions all along the observers' world line.
Thus,
\beq
\mathbf{e}^{ff}_{\hat{\theta}} =  \frac{1}{\tilde{r}}\,\mathbf{e}_{\vit}\,,\,
\mathbf{e}^{ff}_{\hat{\phi}} = \frac{1}{\tilde{r}\sin\vit}\,\mathbf{e}_{\vip}\,.
\label{eq:freevinkorto}
\eeq
Finally, the spacelike unit basis vector associated with the freely falling observer and pointing in the $r$--direction
of the Schwarzschild coordinates may be calculated by writing
$\mathbf{e}^{ff}_{\hat{r}}=a_{ff}^{\ti}\,\mathbf{e}_{\ti}+a_{ff}^{\ra}\,\mathbf{e}_{\ra}$ and then solving the two equations
$\mathbf{e}^{ff}_{\hat{r}}\cdot\mathbf{e}^{ff}_{\hat{\tau}}=0$ and $\mathbf{e}^{ff}_{\hat{r}}\cdot\mathbf{e}^{ff}_{\hat{r}}=1$
for the components $a_{ff}^{\ti}$ and $a_{ff}^{\ra}$. The result is
\beq
a_{ff}^{\ti} = \frac{\frac{R_S}{\ra}\sqrt{\curvconst}-\sqrt{\radvelf}}{\left(\curv\right)c}\,,
\label{eq:freetimecomp}
\eeq
and
\beq
a_{ff}^{\ra} = \sqrt{\curvconst}\,.
\label{eq:freeradcomp}
\eeq
Simple calculations will verify that the set of mutually orthogonal unit basis vectors
$\{\mathbf{e}_{\hat{\tau}}^{ff},\mathbf{e}^{ff}_{\hat{r}},\mathbf{e}^{ff}_{\hat{\theta}},\mathbf{e}^{ff}_{\hat{\phi}}\}$
satisfy the requirements for an orthonormal basis, that is,
\beq
\mathbf{e}^{ff}_{\hat{\mu}}\cdot\mathbf{e}^{ff}_{\hat{\nu}} = \eta_{\hat{\mu}\hat{\nu}}\,,
\label{eq:reqfreeorthobasis}
\eeq
all along the freely falling observer's world line. Note also that in the limit $\ra\rightarrow \ra_0$, these unit basis
vectors become identical to the unit basis vectors in Eqs.~(\ref{eq:statimeorto})--(\ref{eq:staradorto}) associated with the
stationary observer at rest at $\ra=\ra_0$. This identification is a consequence of the freely falling observer having zero
velocity relative to the stationary observer at the initial position.

Now, let $(\tau_{ff},\hat{x}_{ff}^r,\hat{x}_{ff}^\theta,\hat{x}_{ff}^\phi)$ denote the coordinates in the proper reference
frame of the freely falling observer such that the coordinate axes of the rectangular grid
$(\hat{x}_{ff}^r,\hat{x}_{ff}^\theta,\hat{x}_{ff}^\phi$) are aligned with the directions of the orthonormal basis vectors
$\{\mathbf{e}^{ff}_{\hat{r}},\mathbf{e}^{ff}_{\hat{\theta}},\mathbf{e}^{ff}_{\hat{\phi}}\}$. Assume that the freely falling
observer, at some arbitrary radial position $\ra$ of his world line, emits a particle (with nonzero rest mass) having
initial velocity $v_e$ along the $\hat{x}_{ff}^r$--direction. The initial velocity of the particle as observed in the
local laboratory of the freely falling observer may then be written
\beq
\mathbf{v}_e = \frac{d\hat{x}_{ff}^r}{d\tau_{ff}}\,\mathbf{e}^{ff}_{\hat{r}} = v_e\,\mathbf{e}^{ff}_{\hat{r}}
= v_e\left(a_{ff}^{\ti}\,\mathbf{e}_{\ti} + a_{ff}^{\ra}\,\mathbf{e}_{\ra}\right)\,.
\label{eq:freefallinitvel}
\eeq
Here $v_e>0$ for a particle moving in the positive direction along the coordinate axis $\hat{x}_{ff}^r$ and $v_e<0$
for a particle moving in the negative direction along the same axis, and $a_{ff}^{\ti}$ and $a_{ff}^{\ra}$ are
given in Eqs.~(\ref{eq:freetimecomp}) and (\ref{eq:freeradcomp}). To calculate the coordinate velocity of the
particle in Eddington-Finkelstein coordinates, we once again examine the four-velocity. The
four-velocity $\mathbf{u}_{e}$ of the emitted particle can be expressed as
\beq
\mathbf{u}_{e} = \frac{dx_{ff}^{\hat{\mu}}}{d\tau_e}\,\mathbf{e}_{\hat{\mu}}^{ff}
= \left(\mathbf{e}_{\hat{\tau}}^{ff}+\frac{d\hat{x}_{ff}^r}{d\tau_{ff}}\,\mathbf{e}_{\hat{r}}^{ff}\right)
\frac{d\tau_{ff}}{d\tau_e}
= \left(\mathbf{e}_{\hat{\tau}}^{ff}+v_e\,\mathbf{e}_{\hat{r}}^{ff}\right)\frac{d\tau_{ff}}{d\tau_e}\,,
\label{eq:fourvelmatorto}
\eeq
where $\tau_e$ denotes the proper time of the emitted particle. Using Eqs.~(\ref{eq:freetimeorto}),
and (\ref{eq:freefallinitvel}) the four-velocity can be rewritten as
\beq
\mathbf{u}_{e} = \left[\left(\dot{\ti}+v_ea_{ff}^{\ti}\right)\mathbf{e}_{\ti}
+\left(v_ea_{ff}^{\ra}+\dot{\ra}\right)\mathbf{e}_{\ra}\right]\frac{d\tau_{ff}}{d\tau_e}\,.
\label{eq:fourvelmatortorew}
\eeq
In the Eddington-Finkelstein coordinates the four-velocity can be expressed as
\beq
\mathbf{u}_{e} = \frac{dx^{\tilde{\mu}}}{d\tau_e}\,\mathbf{e}_{\tilde{\mu}}
= \left(\mathbf{e}_{\ti}+\frac{d\ra}{d\ti}\,\mathbf{e}_{\ra}\right)\frac{d\ti}{d\tau_e}\,.
\label{eq:fourvelmatedfink}
\eeq
Equating the two expressions in Eqs.~(\ref{eq:fourvelmatortorew}) and (\ref{eq:fourvelmatedfink}), we obtain the two equations
\beq
\left(\dot{\ti}+v_ea_{ff}^{\ti}\right)\frac{d\tau_{ff}}{d\tau_e} = \frac{d\ti}{d\tau_e},\;\ \;
\left(v_ea_{ff}^{\ra}+\dot{\ra}\right)\frac{d\tau_{ff}}{d\tau_e} = \frac{d\ra}{d\ti}\frac{d\ti}{d\tau_e}\,,
\label{eq:coordveleqs}
\eeq
to be solved for the coordinate velocity $d\ra/d\ti$. A short calculation yields the solution
\beq
\left(\frac{d\ra}{d\ti}\right)_e = \frac{\left(\curv\right)\left(v_e\sqrt{\curvconst}-\sqrt{\radvelf}\,c\right)}{
\sqrt{\curvconst}\left(1+\frac{v_e}{c}\frac{R_S}{\tilde{r}}\right)-\sqrt{\radvelf}\left(\frac{R_S}{\tilde{r}}+\frac{v_e}{c}\right)}\,.
\label{eq:coordvelsol}
\eeq
We emphasize here that the solution in Eq.~(\ref{eq:coordvelsol}) corresponds to the \emph{initial value} of the emitted particles'
coordinate velocity at the point $\ra$ of emission.

We are particularly interested in the limit $\ra_0\rightarrow\infty$ corresponding to the situation for which the emitter moves
together with the river of space. Evaluating this limit, we finally arrive at
\beq
\left(\frac{d\ra}{d\ti}\right)_e = \frac{\left(\curv\right)}{
1+\frac{v_e}{c}\frac{R_S}{\ra}-\sqrt{\frac{R_S}{\ra}}\left(\frac{R_S}{\ra}+\frac{v_e}{c}\right)}
\left(v_e-\sqrt{\frac{R_S}{\ra}}c\right)\,.
\label{eq:emitvel}
\eeq
Note that $v_e$ now represents the velocity of the emitted particle relative to the inflowing river of space.
Equation~(\ref{eq:emitvel}) exhibits several interesting properties showing the effect of the river of space upon the
motion of particles. In particular we note the appearance of the last term $v_e-\sqrt{R_S/\ra}\,c$ which, in Galilean terms,
represents the velocity of the emitted particle as measured relative to the stationary observers along the particle's world line
(recall here the result of Eq.~(\ref{eq:Newtonvel})).

Consider now the case where the freely falling observer emits photons, i.e., $v_e=\pm c$. Then, we again recover the results
obtained earlier in Eq.~(\ref{eq:lightcoordvel}). Hence, even if the physical velocity of light is isotropic in the freely
falling observer's reference frame, the coordinate velocity of light in the Eddington-Finkelstein coordinates is anisotropic.
Moreover, the coordinate velocity of light emitted outwards vanishes at $\ra=R_S$. This demonstrates that the fountain
picture of Fig.~\ref{fig:fountain} is misleading. The river of space, having an inward--directed physical velocity
$d\hat{x}_s^r/d\tau_s=-\sqrt{R_S/\ra}\,c=-c$ at $\ra=R_S$, prohibits any outward motion of light at $\ra=R_S$.

Next, we proceed to analyze the effect of the river of space upon the motion of material particles. Firstly, if the material
particle is emitted inwards, i.e., $v_e<0$, then it is clear that $\left(d\ra/d\ti\right)_e<0$ for all $\ra$.
However, if the particle is emitted outwards, i.e. $v_e>0$, inspection of Eq.~(\ref{eq:emitvel}) reveals that the first term is
always positive and that the sign of the coordinate velocity therefore is determined by the last term $v_e-\sqrt{R_S/\ra}\,c$.
Accordingly, if the velocity of the particle emitted outwards is too small compared to the river velocity, that is
$v_e<\sqrt{R_S/\ra}\,c$, the particle is drawn inwards by the river of space. In particular, since the river of space has
velocity $\sqrt{R_S/\ra}\,c\geq c$ in the region $\ra\leq R_S$ and all material particles moving outwards must have $v_e<c$,
the coordinate velocity $\left(d\ra/d\ti\right)_e<0$ both at and inside the black hole horizon. Note that this result applies to
all material particles with arbitrary motion because a particle having a velocity $v<c$ with respect to a certain observer
will, by application of the transformation law for velocities from special relativity, have a velocity less than the speed of light
also with respect to the river of space.

\begin{figure}[tbp]
\begin{center}
\includegraphics[width=10.0cm]{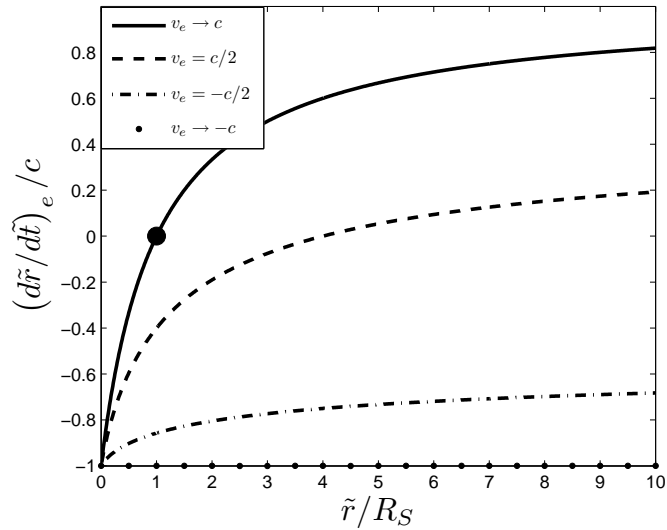}
\caption{Plots of the initial coordinate velocity $\left(d\ra/d\ti\right)_e$ (Eq.~(\ref{eq:emitvel})) as a function of the radial
coordinate $\ra$ for two particles emitted along the positive and negative radial directions in the local laboratory of a freely
falling observer in the Schwarzschild spacetime. Particles emitted radially outward with speeds approaching the speed of light
(solid line) have positive coordinate velocities outside the event horizon and can therefore escape the black hole in this
region. At the horizon (black dot) the coordinate velocity of these particles vanishes. Hence photons emitted radially
outward remain stationary at the horizon. However, inside the horizon the coordinate velocity becomes negative. Thus all
particles are pulled towards the singularity by the river of space inside the horizon. Particles emitted outwards at speeds
smaller than the speed of light (dashed line) are pulled towards the singularity by the river of space also at radii which are
larger than the Schwarzschild radius $R_S$. Particles emitted radially inwards (dash-dotted and dotted lines) have negative
coordinate velocities everywhere and eventually ends up in the singularity at $\ra=0$.}
\label{fig:emitcoordvel}
\end{center}
\end{figure}

Figure~\ref{fig:emitcoordvel} shows the coordinate velocity $\left(d\ra/d\ti\right)_e$ as a function of $\ra$ for four particular
choices of the physical velocity $v_e$ for particles emitted in both directions in the freely falling observer's reference frame.
We can now provide an intuitive interpretation of these peculiar results in light of the river model of space expounded in
Sec.~\ref{subsec:riverinterior}: The freely falling observer who sends out the two particles is, by definition, comoving with the river
of space. The river of space flows inwards toward the singularity at ever increasing speeds with respect to the stationary observers
for decreasing values of the coordinate $\ra$, and flows faster than the speed of light inside the black hole horizon. Hence, even if
a particle is emitted with a certain velocity along the positive radial direction with respect to the infalling observer's laboratory,
the river of space will eventually drag the particle along with it towards smaller $\ra$ as one approaches the black hole horizon.
At and inside the horizon all particles with nonzero rest mass will be pulled towards the singularity by the river of space no matter
how large the particles' outward directed initial velocities $v_e$ $(<c)$ are. Because the river of space flows faster than the
speed of light inside the horizon, even light emitted outwards by an observer inside the horizon will be dragged towards the
central singularity by the river. Light emitted outwards by an infalling observer \emph{at} the Schwarzschild horizon, where space flows
inwards at the speed of light, will hover indefinitely at the constant radius $\ra=R_S$.

Thus the river model of space yields an intuitive illustration of why the 'fountain picture' of light ray paths is misleading.
Radial light rays do \emph{not} become increasingly deflected from their usual straight paths and then bent back towards the
center $\ra=0$, as claimed in the fountain picture of black holes. On the contrary, light rays emitted in the radial direction
continue to move along radial paths. However, light rays emitted radially outwards at the horizon $\ra=R_S$ remain there forever, whereas
light rays emitted radially outwards by an observer inside the horizon immediately fall along the radial direction into the singularity
at $\ra=0$.

The picture of space as a river falling into the black hole, as outlined above, is qualitatively in concordance with the picture
established in "The river model of black holes" that was recently introduced by Hamilton and Lisle~\cite{Hamilton}.

\section{The river of space in the de Sitter spacetime\label{deSitter}}

In this section we consider the motion of a particle moving freely in the de Sitter spacetime as it appears with respect to observers
who are at rest with respect to an arbitrarily chosen central observer. Thus we shall see that space behaves as an inertial river
flowing outwards with a recession velocity corresponding to Hubble's law. Then we consider the effect of the river of space upon
material particles and light rays and show that outside the cosmological event horizon, where the river of space flows radially
outwards at speeds larger than the speed of light, all physical particles will be dragged towards infinity by the river of space.

\subsection{Coordinate velocity of a particle in radial free motion\label{subsec:freemovcoordvel}}

The de Sitter spacetime is the solution to Einstein's field equations for a homogeneous and isotropic space with a positive cosmological
constant\cite{Gron}. In so-called comoving coordinates $\left(\tau,\chi,\theta,\phi\right)$ the de Sitter spacetime with vanishing spatial
curvature is described by the line element
\beq
ds^2 = -c^2d\tau^2 + e^{2H\tau}\left[d\chi^2+\chi^2d\Omega^2\right]\,.
\label{eq:colinedeSit}
\eeq
Here $H$ denotes the Hubble constant and $d\Omega^2=d\theta^2+\sin^2\theta\,d\phi^2$ is the solid angle element. In these coordinates,
the set of constant coordinate positions, $\chi=\mbox{const.}$, defines a set of freely moving reference particles such that the
\emph{physical} distances between the reference particles increase with time. The physical motion of these reference particles defines
the so-called Hubble-flow of an expanding universe. Our aim, however, is to obtain a description of the river of space as it will be
perceived by local observers who remain at rest with respect to the central observer along the freely moving particle's world line. This
will enable us to establish a picture of the river of space in the de Sitter spacetime which strongly resembles the river model of space
that was developed in Sec.\ref{sec:Schwarz} for the Schwarzschild spacetime. For this purpose we shall introduce so-called static
coordinates $\left(t,r,\theta,\phi\right)$ which are more naturally adapted for the description of stationary observers in the de Sitter
spacetime. The static coordinates are related to the comoving coordinates by the coordinate transformations
\beq
\tau = t + \frac{R_H}{2c}\ln\left(1-\left(\frac{r}{R_H}\right)^2\right),\;\ \; \chi = re^{-\frac{c\tau}{R_H}}
= \frac{re^{-\frac{ct}{R_H}}}{\sqrt{1-\left(\frac{r}{R_H}\right)^2}}\,,
\label{eq:stdeSittrans}
\eeq
where $R_H=c/H$ denotes the Hubble distance. By substituting these expressions for $\tau$ and $\chi$ in the line element in
Eq.~(\ref{eq:colinedeSit}), we obtain the form of the line element in static coordinates as
\beq
ds^2 = -\left(1-\left(\frac{r}{R_H}\right)^2\right)c^2dt^2 + \frac{1}{1-\left(\frac{r}{R_H}\right)^2}\,dr^2
+ r^2d\Omega^2\,.
\label{eq:stlinedeSit}
\eeq
This line element shows that the metric components in the coordinate basis of static coordinates are independent
of the time coordinate $t$. The de Sitter solution therefore represents a static spacetime.

It is clear from Eq.~(\ref{eq:stlinedeSit}) that the de Sitter metric expressed in static coordinates is
singular at $r=R_H$ which, as we shall demonstrate below, represents a cosmological event horizon. This is a coordinate
singularity related to the failure of the static coordinates of properly covering the region $r\geq R_H$ and may be removed
by introducing new coordinates. In order to motivate our choice of new coordinates we next consider radial null geodesics
in the static coordinates. Setting $ds^2=d\theta=d\phi=0$, we obtain the equation
\beq
\frac{dt}{dr} = \pm \frac{1}{\left(1-\left(\frac{r}{R_H}\right)^2\right)c}\,.
\label{eq:nullgeodstatdeSit}
\eeq
Our objective is now to construct coordinates which enable us to describe the motion of the river of space as
it crosses from the region $r<R_H$ to the region $r\geq R_H$. Hence we are interested in future--directed
null paths that can pass from the region inside the horizon to the region outside. For that purpose we must choose
the positive sign in Eq.~(\ref{eq:nullgeodstatdeSit}) which corresponds to outward moving light rays.
Integration of Eq.~(\ref{eq:nullgeodstatdeSit}) then yields
\beq
t+\frac{R_H}{2c}\ln\left(\frac{\frac{r}{R_H}-1}{\frac{r}{R_H}+1}\right) = \mbox{const.}
\label{eq:intnullgeod}
\eeq
If we now define a new time coordinate $\ti$ related to the old coordinates by
\beq
\ti = t+\frac{R_H}{2c}\ln\left(\frac{\frac{r}{R_H}-1}{\frac{r}{R_H}+1}\right) + \frac{r}{c}\,,
\label{eq:tcoordtransfdeSit}
\eeq
and combine this definition with Eq.~(\ref{eq:intnullgeod}), it can be seen that outgoing light rays obey the
condition $dr/d\ti=+c$. In other words, by using this time coordinate, outgoing light rays move along $45^{\circ}$
straight lines in a spacetime diagram and thus resembles the motion of light rays in the flat Minkowski
spacetime of special relativity. The three spatial coordinates are unaffected by the transformation to the
new coordinates. For the sake of notational consistency, however, we shall define the new spatial coordinates by
the relation $\ra=r,\,\vit=\theta,\,\vip=\phi$. Then, rewriting the line element in~(\ref{eq:stlinedeSit}) in terms
of the new coordinates, the line element takes the form
\beq
ds^2 = -\left(\hcurv\right)c^2d\ti^2 - 2\hcrosst\, d\ti d\ra + \left(1+\left(\frac{\ra}{R_H}\right)^2\right)d\ra^2
+\ra^2d\tilde{\Omega}^2\,.
\label{eq:extdeSitmet}
\eeq
This metric does not exhibit a singularity at $\ra=R_H$ and is thus suitable for describing the motion of
particles in the entire region $\ra\geq0$.

We now consider a reference particle moving freely along the radial direction, i.e., $d\vit=d\vip=0$. Assume that
the particle starts from rest at the position $\ra=\ra_0$. The Lagrangian of the free particle is
\beq
L = \frac{1}{2}g_{\mu\nu}\dot{x}^{\mu}\dot{x}^{\nu} =
-\frac{1}{2}\left(\hcurv\right)c^2\dot{\ti}^2 - \hcrosst\dot{\ti}\dot{\ra}
+ \frac{1}{2}\left(1+\frac{\ra}{R_H}\right)\dot{\ra}^2\,,
\label{eq:LagrangeHubbleflow}
\eeq
where the dots denote the derivatives with respect to the proper time $\tau_F$ of the free particle. It is
now straightforward to calculate the coordinate velocity by following exactly the same procedure that was used in
Sec.~\ref{subsec:freefallcoordvel}. The Lagrangian is independent of $\ti$, and hence the momentum conjugate to the
time--coordinate,
\beq
p_{\ti} = \frac{\partial L}{\partial\dot{\ti}} = -\left(\hcurv\right)c^2\dot{\ti}-\hcrosst\dot{\ra}\,,
\label{eq:hmomconj}
\eeq
is constant during the motion. The four-velocity identity $g_{\mu\nu}\dot{x}^{\mu}\dot{x}^{\nu}=-c^2$ yields
\beq
p_{\ti}\dot{\ti} - \hcrosst\dot{\ti}\dot{\ra} + \left(1+\left(\frac{\ra}{R_H}\right)^2\right)\dot{\ra}^2 = -c^2\,.
\label{eq:hfourvelid}
\eeq
Using that $\dot{\ra}\left|_{\ra=\ra_0}\right.=0$ in Eqs.~(\ref{eq:hmomconj}) and (\ref{eq:hfourvelid}), we find
\beq
p_{\ti} = -\sqrt{\hcurvconst}\,c^2\,.
\label{eq:hmomconjconst}
\eeq
Eqs.~(\ref{eq:hmomconj}) and (\ref{eq:hfourvelid}) implies that
\beq
-\left(p_{\ti} - \hcrosst\dot{\ra}\right)\frac{p_{\ti}+\hcrosst\dot{\ra}}{\left(\hcurv\right)c^2}
+ \left(1+\left(\frac{\ra}{R_H}\right)^2\right)\dot{\ra}^2 = -c^2\,,
\label{eq:hradeq}
\eeq
with the corresponding solutions
\beq
\dot{\ra} = \frac{\sqrt{\ra^2-\ra_0^2}\,c}{R_H}\,
\label{eq:hsolradvel}
\eeq
and
\beq
\dot{\ti} = \frac{\sqrt{\hcurvconst}-\frac{\ra^2}{R_H^3}\sqrt{\ra^2-\ra_0^2}}{\hcurv}\,.
\label{eq:hsoltimeder}
\eeq
By combining Eqs.~(\ref{eq:hsolradvel}) and (\ref{eq:hsoltimeder}), the coordinate velocity of the free particle
is found to be
\beq
\frac{d\ra}{d\ti} = \frac{\frac{\sqrt{\ra^2-\ra_0^2}}{R_H}\left(\hcurv\right)}{\sqrt{\hcurvconst}
-\frac{\ra^2}{R_H^3}\sqrt{\ra^2-\ra_0^2}}\,c\,.
\label{eq:hcoordvel}
\eeq
We define the river of space by those freely moving particles that start from rest at the origin. The coordinate
velocity of the river of space is hence given by
\beq
\left(\frac{d\ra}{d\ti}\right)_{\mbox{space}} = \lim_{\ra_0\rightarrow 0}\frac{d\ra}{d\ti}
= \frac{\left(1-\left(\frac{\ra}{R_H}\right)^2\right)\frac{\ra}{R_H}}{1-\left(\frac{\ra}{R_H}\right)^3}\,c\,.
\label{eq:hrivervel}
\eeq
Applying L'Hopital's rule to this expression, we find that the coordinate velocity at the cosmological horizon
is $\left(d\ra/d\ti\right)_{\mbox{space}}=(2/3)\,c$.

The coordinate velocity of light rays may now be found by setting $ds^2=0$ in~(\ref{eq:extdeSitmet}), which
yields the equation
\beq
\left(1+\left(\frac{\ra}{R_H}\right)^2\right)\left(\frac{d\ra}{d\ti}\right)^2 - 2\hcrosst\frac{d\ra}{d\ti}
- \left(\hcurv\right)c^2 = 0\,,
\label{eq:lieqcoordveldeSit}
\eeq
with solutions
\beq
\left(\frac{d\ra}{d\ti}\right)_{l+} = c\,,\ \left(\frac{d\ra}{d\ti}\right)_{l-}
= -\frac{\hcurv}{1+\left(\frac{\ra}{R_H}\right)^2}\,c\,
\label{eq:licoordveldeSit}
\eeq
for the outgoing and ingoing light rays, respectively. The coordinate velocities of light at the cosmological horizon are
then $\left(d\ra/d\ti\right)_{l+}\left|_{\ra=R_H}\right.=c$ and $\left(d\ra/d\ti\right)_{l-}\left|_{\ra=R_H}\right.=0$.

\subsection{The physical velocity field of the river of space\label{subsec:physveldeSit}}

In order to calculate the physical velocity field of the river of space in the de Sitter spacetime, we shall first
introduce a set of orthonormal basis vectors which define the local laboratories of stationary observers. Since for
a stationary observer, $d\ra=d\vip=d\vit=0$, we obtain from the line element in~(\ref{eq:extdeSitmet}) that
$d\ti/d\tau_s=1/\sqrt{1-\left(\ra/R_H\right)^2}$, where $\tau_s$ denotes the proper time of the stationary
observer. The stationary observer's timelike unit basis vector is thus defined as
\beq
\mathbf{e}^s_{\hat{\tau}} = \frac{dx^{\tilde{\mu}}}{d\tau_s}\,\mathbf{e}_{\tilde{\mu}}
= \frac{1}{\sqrt{1-\left(\ra/R_H\right)^2}}\,\mathbf{e}_{\ti}\,.
\label{eq:hstatimeorto}
\eeq
The spacelike unit basis vectors pointing in the $\vit$-- and $\vip$--directions are
\beq
\mathbf{e}^s_{\hat{\theta}} =  \frac{1}{\tilde{r}}\,\mathbf{e}_{\vit}\,,\,
\mathbf{e}^s_{\hat{\phi}} = \frac{1}{\tilde{r}\sin\vit}\,\mathbf{e}_{\vip}\,.
\label{eq:hstavinkorto}
\eeq
Finally, the spacelike unit basis vector pointing in the increasing $r$--direction, where $r$ is the radial
coordinate of the \emph{static} coordinates, is
$\mathbf{e}_{\hat{r}}^s=\sqrt{1-\left(r/R_H\right)^2}\,\mathbf{e}_r$. Using the transformation rule
$\mathbf{e}_r=\left(\partial\tilde{x}^{\nu}/\partial r\right)\mathbf{e}_{\tilde{\nu}}$, where
$\left(\partial\tilde{x}^{\nu}/\partial r\right)$ is given by Eq.~(\ref{eq:tcoordtransfdeSit}), we find
\beq
\mathbf{e}_{\hat{r}}^s = -\frac{\left(\frac{\ra}{R_H}\right)^2}{\sqrt{\hcurv}\,c}\,\mathbf{e}_{\ti}
+ \sqrt{\hcurv}\,\mathbf{e}_{\ra}\,.
\label{eq:hstatradorto}
\eeq

The radial velocity field of a free particle which follows the Hubble flow can now be calculated by following
the same line of reasoning as in Sec.~\ref{subsec:riverexterior}. In the coordinates $\left(\ti,\ra,\vit,\vip\right)$
the four-velocity is given by
\beq
\mathbf{u}_F = \frac{dx^{\tilde{\mu}}}{d\tau_F}\,\mathbf{e}_{\tilde{\mu}}
= \left(\mathbf{e}_{\ti}+\frac{d\ra}{d\ti}\,\mathbf{e}_{\ra}\right)\dot{\ti}\,.
\label{eq:hfourvelfree}
\eeq
In the coordinates corresponding to the stationary observer's proper reference frame,
$(\tau_s,\hat{x}_s^r,\hat{x}_s^\theta,\hat{x}_s^\phi)$, the same four-velocity may be written as
\beq
\mathbf{u}_F = \frac{dx_s^{\hat{\mu}}}{d\tau_F}\,\mathbf{e}_{\hat{\mu}}^s
= \left(\mathbf{e}^s_{\hat{\tau}}+\frac{d\hat{x}_s^r}{d\tau_s}\,\mathbf{e}_{\hat{r}}^s\right)\frac{d\tau_s}{d\tau_F}\,.
\label{eq:hfourvelfreeorto}
\eeq
Substituting for $\mathbf{e}^s_{\hat{\tau}}$ and $\mathbf{e}_{\hat{r}}^s$ from equations~(\ref{eq:hstatimeorto})
and (\ref{eq:hstatradorto}) and equating the expressions in~(\ref{eq:hfourvelfree}) and (\ref{eq:hfourvelfreeorto}),
we obtain the following set of equations:
\beq
\dot{\tilde{t}} = \frac{1}{\sqrt{\hcurv}}\left(1-\left(\frac{\ra}{R_H}\right)^2\frac{1}{c}
\frac{d\hat{x}_s^r}{d\tau_{s}}\right)\frac{d\tau_s}{d\tau_{F}},\;\ \;
\frac{d\ra}{d\ti}\,\dot{\ti} = \sqrt{\hcurv}\frac{d\hat{x}_s^r}{d\tau_{s}}\frac{d\tau_s}{d\tau_{F}}\,.
\label{eq:hfourvelequal}
\eeq
Elimination of the term $d\tau_s/d\tau_F$ from these equations yields
\beq
\frac{d\ra}{d\ti} = \frac{\left(\hcurv\right)
\frac{d\hat{x}_s^r}{d\tau_{s}}}{1-\left(\frac{\ra}{R_H}\right)^2\frac{1}{c}
\frac{d\hat{x}_s^r}{d\tau_{s}}}\,,
\label{eq:hgencoordvelsol}
\eeq
or
\beq
\frac{d\hat{x}_s^r}{d\tau_s}
= \frac{\frac{d\ra}{d\ti}}{1-\left(\frac{\ra}{R_H}\right)^2\left(1-\frac{1}{c}\frac{d\ra}{d\ti}\right)}\,.
\label{eq:hgenortovelsol}
\eeq
From Eq.~(\ref{eq:hgencoordvelsol}) it can be seen that if the stationary observer observes light rays, i.e.,
$d\hat{x}_s^r/d\tau_s=\pm c$, one recovers the result obtained earlier in Eq.~(\ref{eq:licoordveldeSit}).
Hence, as in the case of light rays described in the Eddington-Finkelstein coordinates in Schwarzschild spacetime
in Sec.~\ref{subsec:riverexterior}, we see that even though the physical velocity of light is isotropic, the coordinate
velocity of light in the extended coordinates $\left(\ti,\ra,\vit,\vip\right)$ is anisotropic. However, in striking
contrast to what was obtained in the case of the Schwarzschild spacetime, it is now seen that it is the coordinate
velocity of light rays that are \emph{ingoing} in the stationary observer's proper reference frame which vanish
in the limit $\ra\rightarrow R_H$. As will be discussed further in Sec.~\ref{subsec:velemitdeSit}, this behavior of
light rays in the de Sitter spacetime may be attributed to the effect of the river of space, i.e., the Hubble flow,
having an outward--directed velocity $d\hat{x}_s^r/d\tau_s=c$ at $\ra=R_H$ which therefore prohibits any inward
motion of light in the region $\ra\geq R_H$.

\begin{figure}[tbp]
\begin{center}
\includegraphics[width=10.0cm]{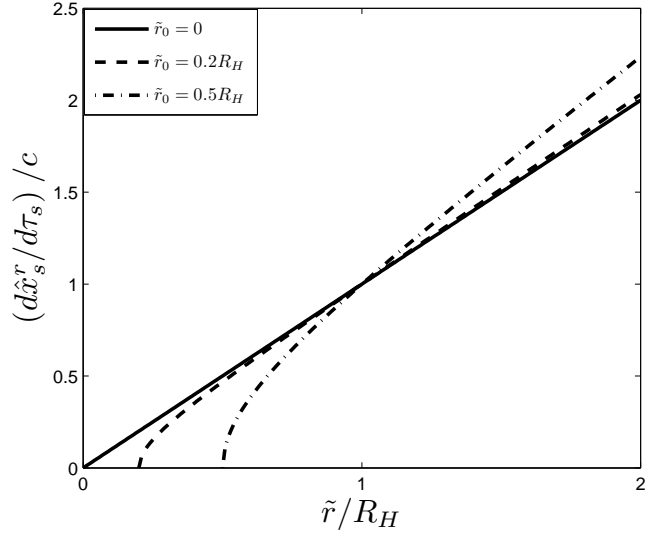}
\caption{Plots of the radial velocity $d\hat{x}_s^r/d\tau_s$ of a freely moving particle in the de Sitter spacetime
as measured by stationary observers in their proper reference frames. The different curves correspond to different
choices of the initial position $\ra_0$ at which the particle is released. Independently of its initial position
$\ra_0$ the particle reaches the speed of light relative to (imaginary) stationary observers at the cosmological
horizon $\ra=R_H$ and continues to increase its speed toward infinity beyond the horizon.}
\label{fig:horthovel}
\end{center}
\end{figure}

Lastly, we obtain the physical velocity field of a freely moving particle as observed by stationary observers
along the particle's world line by substituting for $d\ra/d\ti$ in Eq.~(\ref{eq:hgenortovelsol}) the solution given in
Eq.~(\ref{eq:hcoordvel}), leading to the result
\beq
\frac{d\hat{x}_s^r}{d\tau_s} = \frac{\sqrt{\ra^2-\ra_0^2}}{R_H\sqrt{\hcurvconst}}\,c\,.
\label{eq:hfreepartvelorto}
\eeq
A plot of this velocity field for three different values of $\ra_0$ is shown in Fig.~\ref{fig:horthovel}.
As can be seen from the figure, the speed of a freely moving particle in the de Sitter spacetime approaches the speed
of light $c$ as $\ra\rightarrow R_H$ for all values of the initial position $\ra_0$. Outside the cosmological horizon,
$\ra>R_H$, Eq.~(\ref{eq:hcoordvel}) predicts that free particles move faster than the speed of light with
respect to stationary observers. At and outside the cosmologial horizon, however, real physical observers cannot
be stationary because they must follow timelike world lines in spacetime. Yet, in analogy with the interpretation
of stationary observers at and inside the black hole horizon in Sec.~\ref{subsec:riverinterior}, we may interpret
stationary observers at and outside the cosmological horizon as imaginary tachyon--like observers. The
physical velocity of the river of space in the de Sitter spacetime is
\beq
\left(\frac{d\hat{x}_s^r}{d\tau_s}\right)_{\mbox{space}}
= \lim_{\ra_0\rightarrow 0} \frac{d\hat{x}_s^r}{d\tau_s} = \frac{\ra}{R_H}\,c\,.
\label{eq:hrivervelorto}
\eeq
Thus we conclude that the river of space flows radially outward at speeds smaller than the speed of light inside
the cosmological horizon. At the horizon $\ra=R_H$ the river of space reaches the speed of light and outside the
horizon $\ra>R_H$ the river of space exceeds the speed of light with respect to the imaginary stationary observers.

\subsection{Coordinate velocity of a particle emitted from a freely moving observer\label{subsec:velemitdeSit}}

In order to calculate the coordinate velocity of a particle emitted by a freely moving observer following the Hubble flow,
we first need to construct a set of orthonormal basis vectors representing the observer's local laboratory. The
timelike unit basis vector is
\beq
\mathbf{e}_{\hat{\tau}}^F = \mathbf{u}_F\,,
\label{eq:hunittimevec}
\eeq
where $\mathbf{u}_F$ is the four-velocity of the freely moving observer given in Eq.~(\ref{eq:hfourvelfree}). The spacelike unit
basis vector pointing in the radial direction of the static coordinates can be written
$\mathbf{e}_{\hat{r}}^F=a_F^{\ti}\,\mathbf{e}_{\ti}+a_F^{\ra}\,\mathbf{e}_{\ra}$. By solving the two equations
$\mathbf{e}_{\hat{r}}^F\cdot\mathbf{e}_{\hat{\tau}}^F=0$ and $\mathbf{e}_{\hat{r}}^F\cdot\mathbf{e}_{\hat{r}}^F=1$
for the components $a_F^{\ti}$ and $a_F^{\ra}$, we obtain the results
\beq
a_F^{\ti} = \frac{\frac{\sqrt{\ra^2-\ra_0^2}}{R_H}-\left(\frac{\ra}{R_H}\right)^2\sqrt{\hcurvconst}}
{\left(\hcurv\right)c}\,,
\label{eq:tunitveccomp}
\eeq
and
\beq
a_F^{\ra} = \sqrt{\hcurvconst}\,.
\label{eq:runitveccomp}
\eeq
The remaining two spacelike unit vectors are chosen as
\beq
\mathbf{e}^F_{\hat{\theta}} =  \frac{1}{\tilde{r}}\,\mathbf{e}_{\vit}\,,\,
\mathbf{e}^F_{\hat{\phi}} = \frac{1}{\tilde{r}\sin\vit}\,\mathbf{e}_{\vip}\,.
\label{eq:hfreevinkorto}
\eeq
Next, we let the freely moving observer emit a particle with initial velocity $v_e$ along the radial direction
at some arbitrary position $\ra$ of his world line. If we denote the coordinates in the proper reference frame
of the freely moving observer by $(\tau_{F},\hat{x}_{F}^r,\hat{x}_{F}^\theta,\hat{x}_{F}^\phi)$, then the
initial velocity of the emitted particle can be defined as
\beq
\mathbf{v}_e = \frac{d\hat{x}_F^r}{d\tau_F}\,\mathbf{e}_{\hat{r}}^F = v_e\,\mathbf{e}_{\hat{r}}^F
=v_e\left(a_F^{\ti}\,\mathbf{e}_{\ti}+a_F^{\ra}\,\mathbf{e}_{\ra}\right)\,,
\label{eq:hinitpartvel}
\eeq
where $v_e>0$ for a particle moving in the positive radial direction and $v_e<0$ for a particle moving in the
negative direction, and $a_F^{\ti}$ and $a_F^{\ra}$ are given by Eqs.~(\ref{eq:tunitveccomp}) and
(\ref{eq:runitveccomp}).
In the coordinates of the proper reference frame the four-velocity $\mathbf{u}_e$ of the emitted particle then becomes
\beq
\mathbf{u}_e = \frac{dx_F^{\hat{\mu}}}{d\tau_e}\,\mathbf{e}_{\hat{\mu}}^F
= \left(\mathbf{e}_{\hat{\tau}}^F + v_e\,\mathbf{e}_{\hat{r}}^F\right)\frac{d\tau_F}{d\tau_e}\,,
\label{eq:hfourvelemitorto}
\eeq
where $\tau_e$ denotes the proper time of the emitted particle. Using Eqs.~(\ref{eq:hunittimevec}) and (\ref{eq:hinitpartvel}),
we therefore obtain
\beq
\mathbf{u}_e = \left[\left(\dot{\ti} + v_e a_F^{\ti}\right)\mathbf{e}_{\ti}
+ \left(\dot{\ra} + v_e a_F^{\ra}\right)\mathbf{e}_{\ra}\right]\frac{d\tau_F}{d\tau_e}\,.
\label{eq:hfourvelemitrew}
\eeq
In terms of the global coordinates $\left(\ti,\ra,\vit,\vip\right)$ the same four-velocity is
\beq
\mathbf{u}_e = \frac{dx^{\tilde{\mu}}}{d\tau_e}\mathbf{e}_{\tilde{\mu}}
=\left(\mathbf{e}_{\ti} + \frac{d\ra}{d\ti}\mathbf{e}_{\ra}\right)\frac{d\ti}{d\tau_e}\,.
\label{eq:hfourvelemitglob}
\eeq
By equating the expressions in Eq.~(\ref{eq:hfourvelemitrew}) and Eq.~(\ref{eq:hfourvelemitglob}), we obtain the
two equations to be solved for the coordinate velocity $d\ra/d\ti$:
\beq
\frac{d\ti}{d\tau_e} = \left(\dot{\ti} + v_e a_F^{\ti}\right)\frac{d\tau_F}{d\tau_e}\,,\;\ \;
\frac{d\ra}{d\ti}\frac{d\ti}{d\tau_e} = \left(\dot{\ra} + v_e a_F^{\ra}\right)\frac{d\tau_F}{d\tau_e}\,.
\label{eq:hfourvelequalcoord}
\eeq
Next, we eliminate the term $d\tau_F/d\tau_e$ from these equations and substitute for $\dot{\ti}$ and $\dot{\ra}$
the solutions in~(\ref{eq:hsolradvel}) and (\ref{eq:hsoltimeder}) for a free particle, leading to the solution
\beq
\left(\frac{d\ra}{d\ti}\right)_e
= \frac{\left(\hcurv\right)\left(v_e\sqrt{\hcurvconst}+\frac{\sqrt{\ra^2-\ra_0^2}}{R_H}\,c\right)}
{\sqrt{\hcurvconst}\left(1-\frac{v_e}{c}\left(\frac{\ra}{R_H}\right)^2\right)
+ \frac{\sqrt{\ra^2-\ra_0^2}}{R_H}\left(\frac{v_e}{c}-\left(\frac{\ra}{R_H}\right)^2\right)}\,.
\label{eq:hcoordvelemit}
\eeq
We are particularly interested in the case when the freely moving observer is comoving with the river of
space, that is, when the observer starts from rest at the origin. Accordingly, we evaluate the limit
$\ra_0\rightarrow 0$ to obtain
\beq
\left(\frac{d\ra}{d\ti}\right)_e = \frac{\hcurv}
{1-\frac{v_e}{c}\left(\frac{\ra}{R_H}\right)^2
+ \frac{\ra}{R_H}\left(\frac{v_e}{c}-\left(\frac{\ra}{R_H}\right)^2\right)}\left(v_e+\frac{\ra}{R_H}\,c\right)\,.
\label{eq:hcoordvelemitriver}
\eeq
In this expression $v_e$ is the velocity of the emitted particle relative to the outward flowing river of space.

First we consider the case where the freely moving observer emits light such that $v_e=\pm c$. Then
Eq.~(\ref{eq:hcoordvelemitriver}) reduces to the results given by Eq.~(\ref{eq:licoordveldeSit}). Accordingly,
the isotropic physical velocity of light transforms to become anisotropic in the extended coordinates.
Light rays which the freely falling observer emits inwards in his proper reference frame have vanishing
coordinate velocity in the extended coordinates at the cosmological horizon $\ra=R_H$. In addition, outside
the horizon, $\ra>R_H$, the coordinate velocity of light is always positive. This means that light rays emitted
outside the cosmological horizon must always travel outwards toward infinity and can never enter the interior of
the horizon. The river of space, having an outward-directed motion exceeding the speed of light outside the
horizon, carries everything with it toward infinity.

The effect of the river of space upon material particles can also be made clear by direct inspection of
Eq.~(\ref{eq:hcoordvelemitriver}). Simple analysis reveals that the first term in the expression is always
positive. The sign of the coordinate velocity is therefore determined by the last term $v_e+\left(\ra/R_H\right)c$.
Hence, if the material particle is emitted outwards, corresponding to $v_e>0$, then $\left(d\ra/d\ti\right)_e>0$
for all values of $\ra$. In contrast, when the particle is emitted inwards with $v_e<0$, the sign of the
coordinate velocity is determined by the magnitude of $v_e$ as compared to the magnitude of the velocity
of the river of space. If $\left|v_e\right|>\left(\ra/R_H\right)c$, $\left(d\ra/d\ti\right)_e<0$ and the
particle moves inwards toward smaller $\ra$. But if $\left|v_e\right|<\left(\ra/R_H\right)c$,
$\left(d\ra/d\ti\right)_e>0$ and the particle is drawn outwards by the river of space because its velocity
compared to the river velocity is too small. We emphasize that this is always true in the region $\ra\geq R_H$
because then the river of space has velocity $\left(\ra/R_H\right)c\geq c$ whereas all physical material
particles emitted inwards must obey $\left|v_e\right|<c$.

\begin{figure}[tbp]
\begin{center}
\includegraphics[width=10.0cm]{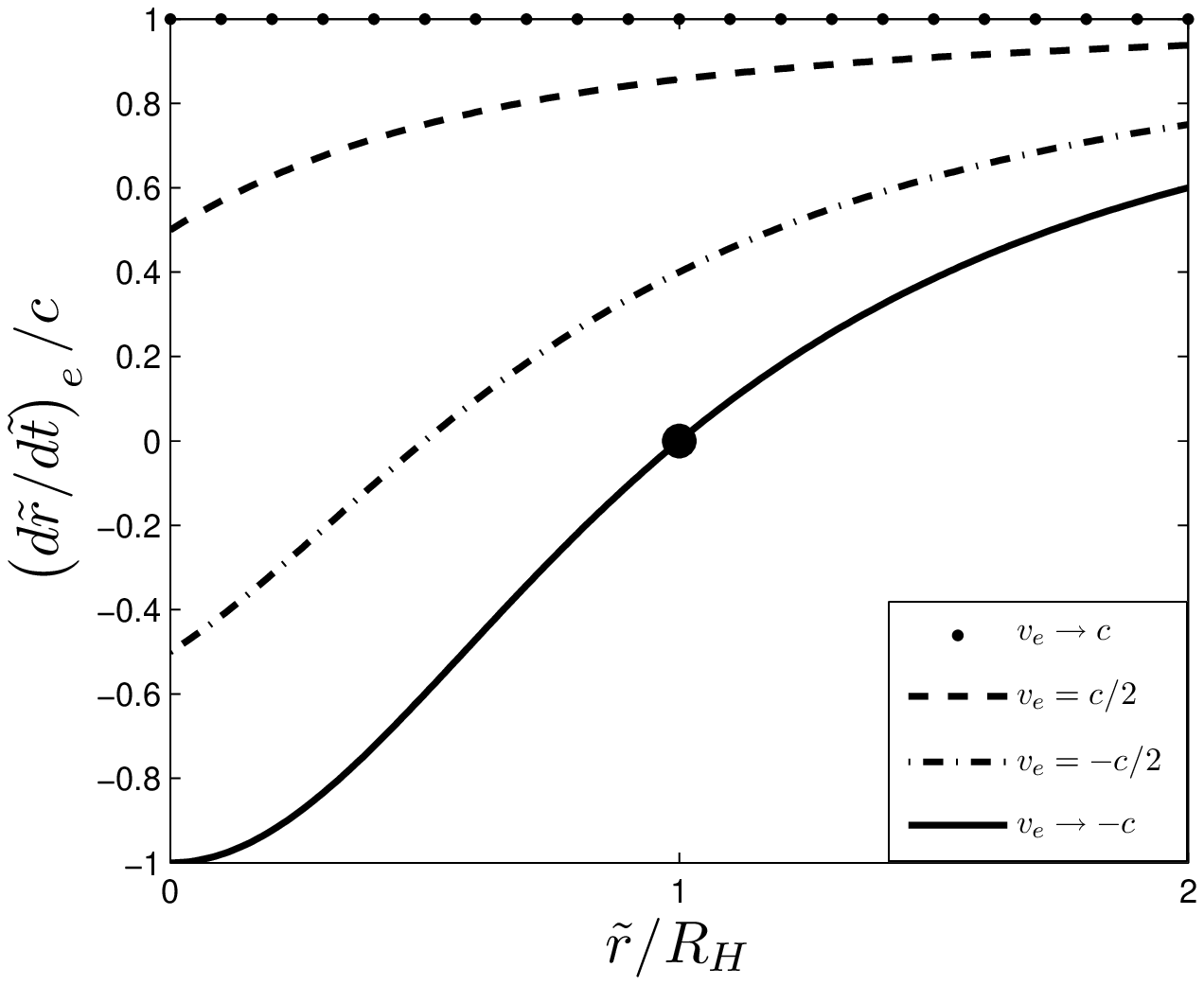}
\caption{Plots of the initial coordinate velocity $\left(d\ra/d\ti\right)_e$ (Eq.~(\ref{eq:hcoordvelemitriver})) as
a function of the radial coordinate $\ra$ for two particles emitted in the positive and negative radial directions
in the local laboratory of a freely moving observer in the de Sitter spacetime. Particles emitted radially inwards
with speeds approaching the speed of light (solid line) have negative coordinate velocities inside the event horizon
and can therefore escape the Hubble flow in this region and reach the observer at the origin. At the horizon (black dot)
the coordinate velocitites of these particles vanish. Hence photons emitted radially inwards remain stationary at the
horizon. However, outside the horizon the coordinate velocity becomes positive. Hence all particles are dragged toward
infinity by the river of space outside the horizon. Particles emitted inwards at speeds smaller than the speed of light
(dash-dotted line) are pulled towards infinity by the river of space at radii which are smaller than the radius of the
cosmological horizon at $\ra=R_H$. Particles emitted radially outwards (dashed and dotted lines) have positive
coordinate velocities everywhere and eventually end up at infinity.}
\label{fig:hemitcoordvel}
\end{center}
\end{figure}

Figure~\ref{fig:hemitcoordvel} illustrates a few examples of the coordinate velocity $\left(d\ra/d\ti\right)_e$
as a function of $\ra$ for different values of the physical velocity $v_e$.
As can be seen, particles emitted in the positive radial direction with respect to the freely moving observer's
laboratory have positive coordinate velocities in the entire region $\ra\geq 0$ and therefore move towards larger
radii with increasing time. In contrast, for small enough values of $\ra$, particles emitted in the negative radial
direction with respect to the freely moving observer's laboratory have negative coordinate velocities and thus move
towards smaller $\ra$. However, as one approaches the cosmological horizon $\ra=R_H$ from below, it is seen that
the coordinate velocity of particles emitted inwards in the local laboratory eventually changes sign and becomes
positive. This change of sign in the coordinate velocity arises from the effects of the river of space: The river of space
flows outwards at ever increasing speeds as $\ra$ increases. The river of space reaches the speed of light with respect
to (imaginary) stationary observers at the cosmological horizon, and continues to exceed the speed of light beyond
the horizon. At and outside the horizon all particles with nonzero rest mass are therefore pulled towards infinity by
the river of space regardless of how large the particles' inward directed velocities $v_e$ are (assuming,
of course, that $\left|v_e\right|<c$). In the particular case of `ingoing' massless photons, however, the coordinate
velocity is seen to vanish at $\ra=R_H$. Thus `ingoing' light rays will remain at rest indefinitely at the horizon,
but outside the horizon even light will be carried toward infinity by the river of space.

The velocity of the river of space given by Eq.~(\ref{eq:hrivervelorto}) also defines the Hubble flow of the
de Sitter universe. To see the connection, use the transformation in~(\ref{eq:stdeSittrans}) to obtain the
velocity expressed in comoving coordinates as $\left(d\hat{x}_s^r/d\tau_s\right)_{\mbox{space}}=He^{H\tau}\chi$
(recall here that $R_H=c/H$). Recognizing $e^{H\tau}$ as the scale factor $a\left(\tau\right)$ and
$a\left(\tau\right)\chi$ as the instantaneous physical distance $l\left(\tau\right)$, we see that
$\left(d\hat{x}_s^r/d\tau_s\right)_{\mbox{space}}=Hl\left(\tau\right)$. Accordingly, the river of space flows
radially outward with a recession velocity corresponding to Hubble's law.

The question of how one should interpret the expansion of the universe has recently been discussed in several
papers (see e.g. Ref.~\cite{Elgaroy} and references therein). We can now address this issue within the context
of the river model of space in the de Sitter spacetime. The freely moving reference particles that constitute
the river of space also represent the fundamental particles that define the three-space in the de Sitter
universe, i.e., the coordinates $\left(\chi,\theta,\phi\right)$ that enter into the spatial part of the line
element~(\ref{eq:colinedeSit}) are comoving with these reference particles. Then, since the river flows
radially outward relative to the rigid frame of reference defined by the family of observers who remain
at rest with respect to the observer at the origin, it is clear that the comoving galaxies will be observed
by the central observer to be moving apart in this universe. Because the origin of the \emph{static}
coordinates is arbitrarily chosen, every other such comoving central observer will observe exactly the same
recession of galaxies relative to the rigid reference frames associated with themselves.
However, as is evident from Eq.~(\ref{eq:hrivervel}), galaxies which are comoving with the river of space
continue their motion towards larger radii even beyond the cosmological horizon at $\ra=R_H$. But, as
was clarified in Sec.~\ref{subsec:velemitdeSit}, the rigid reference frame associated with a comoving
observer \emph{physically} ceases to exist at and beyond the cosmological horizon. All real physical
observers are pulled towards infinity by the river of space beyond the horizon. Thus, a notion
of space where the galaxies physically move \emph{through} the three-space associated with the rigid
reference frame of a comoving observer is inapplicable to the entire three-space of the universe.
We also note that the world lines of the stationary observers connected with the rigid reference frame
deviate from the preferred set of curves discussed in the introduction.

Hence, according to this analysis, it seems natural to literally define the physical three-space
of the universe by the freely moving reference particles. In this way one obtains a description of the
physical three-space of the universe which is globally consistent. With this global general relativistic
interpretation of space it is clear that galaxies move apart because the space between them expands and
not because they move through space. Gr\o n and Elgar\o y~\cite{Elgaroy} have argued for a similar
interpretation of space through an analysis of the geodesic equation in Friedmann universe models and
the empty Milne model.

\section{Conclusion}

In this work we have shown how the concept of global space in general relativity can be conceived of as a continuum of free
reference particles that represent the local inertial frames which define space locally. Within this pictorial framework space
behaves like a river which flows relative to the stationary observers according to the velocity fields given by
Eqs.~(\ref{eq:Newtonvel}) and (\ref{eq:hrivervelorto}) in the Schwarzschild- and the de Sitter spacetimes, respectively.
The river model of space thus provides a vivid picture of the dynamical nature of space as it appears in these spacetimes.

In the Schwarzschild spacetime the river of space flows radially inwards from infinity and into the black hole singularity.
At the horizon of the black hole the river of space reaches the speed of light relative to imaginary stationary observers.
Hence, light rays moving outward relative to the river of space remain stationary at the horizon. This provides an
intuitive explanation as to why the fountain picture in Fig.~\ref{fig:fountain} is misleading. Inside the horizon of
the black hole the river of space flows faster than the speed of light relative to imaginary stationary observers.
Thus, light rays cannot escape from the interior of a black hole, but are pulled towards the singularity by the river of
space. Moreover, since physical particles with nonzero rest mass have velocities which are strictly smaller than the
speed of light, all material particles both at and inside the horizon will be carried towards the singularity
by the infalling river of space. The intuitive picture established by the river model of space in the Schwarzschild
spacetime qualitatively concurs with `The river model of black holes' introduced in an earlier work by
Hamilton and Lisle~\cite{Hamilton}.

The river of space in the de Sitter spacetime flows radially outwards from the origin of an arbitrarily chosen
comoving observer with a recession velocity corresponding to Hubble's law. At the cosmological horizon associated with
the central observer the river of space reaches the speed of light relative to imaginary stationary observers and,
outside the cosmological horizon, the river of space flows radially outwards faster than the speed of light. Light
rays emitted inwards relative to the river of space therefore remain stationary at the cosmological horizon and
are pulled towards infinity by the river of space outside the horizon. Similarly, all material particles with nonzero
rest mass released either at or outside the horizon are drawn towards infinity by the river of space.

Finally, we have shown that the river model of space in the de Sitter spacetime leads to a natural interpretation of
the expansion of space. At and outside the cosmological horizon the effect of the river of space upon material
particles implies that the very notion of a physical rigid reference frame ceases to exist. Hence, regarding the
\emph{global} three-space of the universe, the notion that galaxies literally move through three-space is not
consistent. In fact, a globally consistent interpretation would be the opposite view: that the stationary observers
of the rigid reference frame move through the global three-space of the universe. Thus we arrive at the conclusion
that the relative `motion' of the galaxies must be attributed to an expansion of the three-space between them.

\appendix

\section{The three-space of observers comoving with the river of space\label{appendix}}

An observer falling freely from rest infinitely far away from the central mass is comoving with the river of space.
He has a constant comoving radial coordinate $\rho$ and carries with him a standard clock showing proper
time $\tau$. The transformation between the comoving coordinates $\left(\tau,\rho\right)$ and the Schwarzschild
coordinates $\left(t,r\right)$ have been deduced in Refs.~\cite{Landau,Lightman} as well as by S. Dai and
C. B. Guan~\cite{Dai}:
\beq
t = \tau + R_S\left(\ln\frac{\sqrt{\frac{r}{R_S}}+1}{\sqrt{\frac{r}{R_S}}-1}-2\sqrt{\frac{r}{R_S}}\right),\;\ \;
r^{3/2} = -\frac{3}{2}\sqrt{R_S}\left(\tau+\rho\right)\,.
\label{eq:Schwarzcomovtransf}
\eeq
In terms of the comoving coordinates in the river of space the line element of the Schwarzschild spacetime
takes the form
\beq
ds^2 = -d\tau^2 + \left(\frac{2}{3}\frac{R_S}{\tau+\rho}\right)^{2/3}d\rho^2 + \left(\frac{3}{2}\sqrt{R_S}
\left(\tau+\rho\right)\right)^{4/3}d\Omega^2\,,
\label{eq:Schwarzcomovlineel}
\eeq
or
\beq
ds^2 = -d\tau^2 + \frac{R_S}{r^2}\,d\rho^2 + r^2d\Omega^2\,.
\label{eq:Schwarzmixlineel}
\eeq
This corresponds to the line element of an inhomogeneous universe with anisotropic and position dependent
scale factor.

The simultaneity space of observers moving with the river of space is given by
\beq
dl^2 = \left(\frac{2}{3}\frac{R_S}{\tau+\rho}\right)^{2/3}d\rho^2 + \left(\frac{3}{2}\sqrt{R_S}
\left(\tau+\rho\right)\right)^{4/3}d\Omega^2\,.
\label{eq:Schwarzsimspace}
\eeq
An observer with $\rho=\rho_0$ has initially a large negative value of $\tau$ which increases towards
$-\rho_0-\left(2/3\right)R_S$ as the observer passes the Schwarzschild horizon. Hence three-space expands
in the radial directon and contracts in the tangential direction and, as the observer passes the Schwarzschild
horizon, the spatial line element reduces to the Euclidean form
\beq
dl^2 = d\rho^2 + R_S^2\,d\Omega^2\,.
\label{eq:Schwarzhorizlinel}
\eeq
There is no coordinate singularity at the Schwarzschild horizon with these coordinates. The observer arrives
at the physical singularity at $r=0$ at a proper time $\tau_0=-\rho_0$. The river of space sinks into this
singularity.

In order to calculate the coordinate velocity in the Schwarzschild coordinates of a point with constant
comoving coordinate, $\rho=\mbox{const.}$, we first differentiate the expression for $r$ in
Eq.~(\ref{eq:Schwarzcomovtransf}) with respect to $\tau$, giving
\beq
dr = -\sqrt{\frac{R_S}{r}}\,d\tau\,.
\label{eq:diffradial}
\eeq
Then we substitute the expression for $r$ into that for $t$ in Eq.~(\ref{eq:Schwarzcomovtransf}) and
differentiate with respect to $\tau$. This leads to
\beq
dt = d\tau - \frac{\sqrt{R_S/r}}{1-\frac{R_S}{r}}\,dr\,.
\label{eq:difftime}
\eeq
From Eqs.~(\ref{eq:diffradial}) and (\ref{eq:difftime}) it follows that the coordinate velocity is,
\beq
\frac{dr}{dt} = -\left(1-\frac{R_S}{r}\right)\sqrt{\frac{R_S}{r}}\,,
\label{eq:coordvelapp}
\eeq
in agreement with the velocity of a free particle falling from infinity as calculated from the geodesic
equation in Schwarzschild coordinates~\cite[Eq. 9.39]{Hartle}. In Schwarzschild coordinates the velocity
decreases to zero at the horizon due to the gravitational time dilation.


\begin{thebibliography}{99}
\bibitem{Wald} R. Wald, \emph{General Relativity}, The University of Chicago Press, Chicago (1984).
\bibitem{Gron} \O. Gr\o n and S. Hervik, \emph{Einstein's general theory of relativity}, Springer, New York (2007).
\bibitem{Kaufmann} W. J. Kaufmann, \emph{Black holes and warped spacetime}, W. H. Freeman and Company, San Francisco (1979).
\bibitem{Hamilton} A. J. S. Hamilton and J. P. Lisle, "The river model of black holes", Am. J. Phys. \textbf{76}, 519 (2008).
\bibitem{Hartle} J. B. Hartle, \emph{Gravity: an introduction to Einstein's general relativity}, Addison-Wesley, San Francisco (2003).
\bibitem{Muller} T. M\"{u}ller, "Falling into a Schwarzschild black hole", Gen. Relativ. Gravit. \textbf{40}, 2185 (2008).
\bibitem{MTW} C. W. Misner, K. S. Thorne and J. A. Wheeler, \emph   {Gravitation}, W. H. Freeman, San Francisco (1970), pp. 327-332.
\bibitem{Cavalleri73} G. Cavalleri and G. Spinelli, "Motion of particles entering a Schwarzschild field",
Lett. Nuovo Cimento \textbf{6}, 5 (1973).
\bibitem{Cavalleri77} G. Cavalleri and G. Spinelli, "Note on motion in the Schwarzschild field", Phys. Rev. D \textbf{15}, 3065 (1977).
\bibitem{Cavalleri78} G. Cavalleri and G. Spinelli, "Particle speed when approaching the Schwarzschild radius",
Lett. Nuovo Cimento \textbf{22}, 113 (1978).
\bibitem{Janis73} A. I. Janis, "Note on motion in the Schwarzschild field", Phys. Rev. D \textbf{8}, 2360 (1973).
\bibitem{Janis77} A. I. Janis, "Motion in the Schwarzschild field: A reply", Phys. Rev. D \textbf{15}, 3068 (1977).
\bibitem{Crawford} P. Crawford and I. Tereno, "Generalized observers and velocity measurements in general relativity",
Gen. Rel. Gravit. \textbf{34}, 2075 (2002).
\bibitem{Jaffe} J. Jaffe and I. I. Shapiro, "Lightlike behavior of particles in a Schwarzschild field", Phys. Rev. D \textbf{6}, 405 (1972).
\bibitem{Sudarshan} O. M. P. Bilaniuk, V. K. Deshpande and E. C. G. Sudarshan, ""Meta" Relativity", Am. J. Phys. \textbf{30}, 718 (1962).
\bibitem{Feinberg} G. Feinberg, "Possibility of faster-than-light particles", Phys. Rev. \textbf{159}, 1089 (1967).
\bibitem{Treumann} R. A. Treumann, "Radiation from transcendent matter", Europhys. Lett. \textbf{16}, 121 (1991).
\bibitem{Elgaroy} \O. Gr\o n and \O. Elgar\o y, "Is space expanding in the Friedmann universe models?",
Am. J. Phys. \textbf{75}, 151 (2007).
\bibitem{Landau} L. Landau and E. M. Lifshitz, \emph{The Classial Theory of Fields}, 4th ed., Reed Educational and Professional
Publishing Ltd, Oxford (2002).
\bibitem{Lightman} A. P. Lightman, W. H. Press, R. H. Price and S. A. Teukolsky, \emph{Problem book in relativity and gravitation},
Princeton University Press, Princeton, New Jersey (1975).
\bibitem{Dai} S. Dai and C. B. Guan, "Maximally symmetric subspace decomposition of the Schwarzschild black hole",
arXiv:gr-qc/0406109v1 (2004).
\end{thebibliography}
\end{document}